\documentclass[journal]{IEEEtran}
\usepackage{amssymb}
\usepackage{amsfonts}
\usepackage{amsmath}
\usepackage{algorithm}
\usepackage{algorithmic}
\usepackage{cite}
\usepackage{stfloats}
\usepackage{amsmath}
\usepackage{graphicx}
\usepackage{amsfonts,amsmath,amssymb}
\usepackage{url}
\usepackage{bm}
\usepackage{bbm}
\usepackage{subfigure}
\usepackage{algorithmic}
\usepackage{algorithm}
\usepackage{color}
\newcommand{\Hnull}{\mathcal{H}_0}
\newcommand{\Halt}{\mathcal{H}_1}

\newcommand{\Honull}{\mathcal{{D}}_0}
\newcommand{\Hoalt}{\mathcal{{D}}_1}

\newtheorem{theorem}{\textbf{Theorem}}

\newtheorem{lemma}{\textbf{Lemma}}

\usepackage{amssymb}
\usepackage{}
\usepackage{amsfonts}
\usepackage{amsmath}
\usepackage{algorithm}
\usepackage{algorithmic}
\usepackage{cite}

\usepackage{graphicx,subfigure,amsmath,amssymb,cite}
\usepackage{float}
\normalsize
\ifCLASSINFOpdf
\else
\fi
\begin{document}
\title{{Intelligent Reflecting Surface (IRS)-Aided Covert Wireless Communications with Delay Constraint}}
\author{Xiaobo Zhou, \IEEEmembership{Member, IEEE,} Shihao Yan, \IEEEmembership{Member, IEEE,} Qingqing Wu, \IEEEmembership{Member, IEEE,}  \\Feng Shu, \IEEEmembership{Member, IEEE,} and Derrick Wing Kwan Ng, \IEEEmembership{Fellow, IEEE}


\thanks{X. Zhou is with the Institute of Intelligent Agriculture, School of Information and Computer, Anhui Agricultural University, Hefei 230036, China, and also with the School of Physics and Electronic Engineering, Fuyang Normal University, Fuyang 236037, China (e-mail: zxb@fynu.edu.cn).}

\thanks{S. Yan and D. W. K. Ng are with the School of Electrical Engineering and Telecommunications, University of New South Wales, Sydney, 2052, Australia (email: shihao.yan@unsw.edu.au; w.k.ng@unsw.edu.au).}

\thanks{Q. Wu is with the State Key Laboratory of Internet of Things for Smart City and Department of Electrical and Computer Engineering, University of Macau, Macao, 999078, China (e-mail:qingqingwu@um.edu.mo).}

\thanks{F. Shu is with the School of Information and Communication Engineering, Hainan University, Hainan 570228, China (e-mail: shufeng0101@163.com).}

}

\maketitle

\begin{abstract}
This work examines the performance gain achieved by deploying an intelligent reflecting surface (IRS) in covert communications. To this end, we formulate the joint design of the transmit power and the IRS reflection coefficients by taking into account the communication covertness for the cases with global channel state information (CSI) and without a warden's instantaneous CSI.
For the case of global CSI, we first prove that perfect covertness is achievable with the aid of the IRS even for a single-antenna transmitter, which is impossible without an IRS. Then, we develop a penalty successive convex approximation (PSCA) algorithm to tackle the design problem. Considering the high complexity of the PSCA algorithm, we further propose a low-complexity two-stage algorithm, where analytical expressions for the transmit power and the IRS's reflection coefficients are derived. For the case without the warden's instantaneous CSI, we first derive the covertness constraint analytically facilitating the optimal phase shift design. Then, we consider three hardware-related constraints on the IRS's reflection amplitudes and determine their optimal designs together with the optimal transmit power. Our examination shows that significant performance gain can be achieved by deploying an IRS into covert communications.
\end{abstract}

\IEEEpeerreviewmaketitle

\section{Introduction}
To meet the ever-increasing demand for high-data rate applications and massive connections in wireless networks, multiple advanced technologies, such as massive multiple-input multiple-out (MIMO), millimeter wave (mmWave), and ultra-dense network (UDN), have been advocated\cite{Key20175G,Zhang2020Prospective}. However, these technologies generally suffer from high energy consumption or high hardware complexity, due to the use of a large number of power-hungry radio frequency (RF) chains. As a remedy,
intelligent reflecting surface (IRS) is emerging as a promising solution to improving the spectral and energy efficiency effectively\cite{Wu2021IRStutorial}. Specifically, IRS is a planar surface consisting of a large number of re-configurable and low-cost passive reflecting elements, each of which is able to reflect the incident signals with controllable amplitudes and phase shifts. Thus, IRS can customize the propagation environment from the transmitter to receiver to achieve various design objectives (e.g., signal enhancement, interference suppression).
Due to the aforementioned advantages, IRS has been investigated in various application scenarios, e.g., single-user systems\cite{Ning2020Beamforming,Han2019Large}, multi-user systems\cite{Peng2021AnalysisRis,Zhi2021PowerSca,Wu2019Intelligent1,Huang2019Reconfigurable,Pan2020Multicell,Huang2020Jsac}, and wireless information and power transfer systems\cite{Pan2020Intelligent1,Wu2020Weighted},
and it has been considered as a promising technology for enabling the sixth-generation (6G) wireless networks \cite{Pan2021Reconfigurable6G,Huang2020Holog}.

Recently, considering the increasing concerns on security issues in wireless communications,
several recent works addressed communication security in the context of IRS-assisted wireless networks from the perspective of physical layer security, e.g., \cite{Cui2019Secure,Chu2020Intelligent,Guan2020Intelligent,Dong2020Enhancing,Dong2020Secure,Artificial2020Hong,Yu2020Robust,Hong2020Robust}. In general, the secrecy performance of IRS-assisted networks can be improved by properly designing the IRS reflection coefficients to simultaneously enhance the received signal strength at desired users and weaken them at eavesdroppers.
For example, as shown in \cite{Cui2019Secure}, by jointly optimizing the transmit beamforming together with the reflect beamforming, physical layer security is guaranteed in IRS-assisted network, even if the eavesdropping channel quality is higher than that of the legitimate channel. Along this direction, an alternative optimization algorithm based on semidefinite programming (SDP) relaxation technique was proposed to determine the secure transmit beamforming and reflecting phase shifts in\cite{Chu2020Intelligent}.
In addition, the authors of \cite{Guan2020Intelligent} tackled the question
whether and when artificial noise (AN) is beneficial to the physical layer security in IRS-assisted wireless communication systems.
Meanwhile, MIMO wiretap channels were considered in \cite{Dong2020Enhancing,Dong2020Secure,Artificial2020Hong}
for optimizing the transmit covariance matrix and IRS phase shifts, and the channel imperfectness on multiuser multiple-input-single-output (MISO) and MIMO wireless secure communications were considered in \cite{Yu2020Robust} and \cite{Hong2020Robust}, respectively.

The aforementioned physical layer security technologies focus on protecting the content of the transmitted message against eavesdropping. However, these technologies cannot alleviate privacy issues posed by discovering the presence of the transmitter or transmissions. Fortunately, the emerging and cutting-edge covert communication technology, which aims at hiding the existence of a wireless transmission, is able to preserve such a high-level security and privacy\cite{Yan2019Low}. In general, a positive covert transmission rate can be achieved when the warden (Willie) has various uncertainties, e.g., noise uncertainty\cite{BiaoHe2017on} and channel uncertainty\cite{Wang2019Covert}. In particular,
the fundamental limits of covert communication in additive white Gaussian
noise (AWGN) channels was established in \cite{Bash2013Limits}, where the authors proved that at most $\mathcal{O}(\sqrt{n})$ bits of information can be covertly and reliably conveyed to from a transmitter (Alice) to a desired receiver (Bob) over $n$ channel uses.
In addition, covert communication with the help of a full duplex (FD) receiver was examined in \cite{Shahzad2018Achieving}, where a FD receiver generates AN with a random transmit power to deliberately confuse Willie's detection.
Inspired by this, the authors of \cite{Li2020Optimal} introduced an uninformed jammer to impose artificial uncertainty on Willie. It was revealed that, under the average covertness constraint, the optimal transmit power strategy is in the form of truncated channel inversion.
Meanwhile, covert communication in relaying networks and unmanned aerial vehicle (UAV) networks was examined in \cite{Hu2019Covert} and \cite{Zhou2019Joint,Wang2020Secrecy}, respectively.
Furthermore, the conditions for guaranteeing the optimality of Gaussian signalling for covert communication was addressed in \cite{yan2018gaussian}. 
Most recently, covert communication in random wireless networks and covert communication with delay constraints were investigated in \cite{HeB2018Covert,Zheng2019Multi} and \cite{Shihao2018Delay}, respectively.

Although the aforementioned works on covert communication, i.e., \cite{Yan2019Low,BiaoHe2017on,Wang2019Covert,Bash2013Limits,Shahzad2018Achieving,Li2020Optimal,Hu2019Covert,Zhou2019Joint,Wang2020Secrecy,yan2018gaussian,HeB2018Covert,Zheng2019Multi,Shihao2018Delay}, have studied various strategies to improve its performance, the achievable covert communication rate is still in a low regime due to the stringent covertness requirement (e.g., a low transmit power).
We note that the IRS has the capability of simultaneously enhancing the received signals at a legitimate receiver and deteriorating them at a warden.
{Therefore, the IRS technique is practically appealing in improving covert communication performance, which has been pointed out in a recently published magazine article~\cite{Lu2020Intelligent}.
 In addition, the impact of a warden's noise uncertainty on IRS-assisted covert communication was examined under the assumption of infinite number of channel uses (i.e., without delay constraints) in \cite{Zan2020Covert}, where the reflection beamforming was optimized with fixed IRS reflection amplitudes. We note that the communication delay from a transmitter to a receiver generally increases as the number of channel uses increases. In order to meet the requirement in some low-latency applications (e.g., real-time video processing, connected vehicles), a short packet (i.e., a finite blocklength) should be considered.
We also note that the finite blocklength implies that the transmission should occur within the available channel uses, which is essentially a delay constraint. In addition, covert communications with a finite blocklength is fundamentally different from that with an infinite blocklength, which is due to that the decoding error probability is not negligible in short-packet communications. As such, the main challenge of the covert communications with a finite blocklength arises from that the coding strategy needs to balance between the receiver's decoding error probability and the warden's detection error rate.
 Furthermore, in the context of covert communication, fixing the IRS reflection amplitudes to $1$ may limit its performance and thus may not be optimal.} Against this background, this work considers the delay-constrained IRS-assisted covert communication, where Alice wants to transmit information to Bob covertly in a finite blocklength with the aid of an IRS, while Willie intends to detect the existence of this transmission. We jointly design Alice's transmit power, the IRS reflection amplitudes and phase shifts to enhance covert communication performance. The main contributions of this work are summarized as below.

\begin{itemize}
\item Considering global CSI being available, we prove that perfect covertness (i.e., Willie's detection is equivalent to a random guess) with non-zero transmit power is achievable for a single-antenna Alice in the IRS-assisted covert communication system. Specifically, our analysis reveals that the condition for achieving the perfect communication covertness is that the quality of the channel of Alice-IRS-Willie is better than that of the channel Alice-Willie. Intuitively, this is due to that the reflected signals from the IRS is able to cancel the signals transmitted from Alice directly to Willie. It is found that without the deployment of an IRS, it is impossible to achieve such perfect covertness for a single-antenna Alice, due to the lack of a null space in the channel from Alice to Willie. This result demonstrates the importance of IRS for covert communication.


\item 
    With the global CSI, we first prove that the covertness constraint is convex. We then transform the optimization problem into a generalized nonlinear convex programming (GNCP) and develop a penalty successive convex approximation (PSCA) algorithm to jointly optimize Alice's transmit power together with the reflection amplitudes and phase shifts at the IRS. In order to reduce the computational complexity, we further develop a low-complexity two-stage algorithm, where we derive analytical expressions for Alice's transmit power and the IRS's reflection coefficients.

\item  Considering the case without Willie's instantaneous CSI, we first prove that the Kullback-Leibler (KL) divergence adopted in the covertness constraint is a monotonically increasing function of the received power at Willie, based on which we derive a closed-form expression for the covertness constraint in this case. Our analysis reveals that the covertness constraint is independent of the phase shift of each IRS element, which only affects the signal-to-noise ratio (SNR) at Bob. This observation facilitates the design of the optimal IRS phase shifts. Then, we consider three different practical constraints on the IRS's reflection amplitudes $\rho_n$, i.e., $\rho_n=1, \forall n$, $\rho_n=\rho_0, \forall n$, and $0\leq \rho_n\leq 1, \forall n$, under which the IRS's reflection amplitudes and Alice's transmit power are determined. Our examination shows that the considered IRS-assisted system can significantly outperform the system without an IRS in the context of covert communications in both the considered CSI scenarios.

%
%
%
\end{itemize}

The remainder of this work is organized as follows. Section
II presents the considered system model. Section III and Section IV respectively present the covert communication design for the cases with global CSI and without Willie's instantaneous CSI. Section V provides our numerical results, where the impact of the IRS's location is also examined. The paper is concluded in Section VI.

$\emph{Notation}$: Vectors and matrices are denoted by Boldface lowercase and uppercase letters, respectively. $\mathbf{A}^T$, $\mathbf{A}^H$, represent transpose, conjugate transpose, respectively, while $\mathbf{A}\succeq\mathbf{0}$  denotes semidefiniteness of matrix $\mathrm{\mathbf{A}}$. $\mathbb{E}_x[\cdot]$, $\arg(\cdot)$, $\|\cdot\|$, $\|\cdot\|_1$, and $|\cdot|$ denote the statistical expectation of $x$, phase, $\ell_2$-norm, $\ell_1$-norm, and the absolute value, respectively.
$\mathrm{Pr}\{\cdot\}$ and $\mathcal{CN}(\mu,\sigma^2)$ denote the probability of an event and Gaussian distribution with mean $\mu$ and variance $\sigma^2$, respectively.

\begin{figure}[!t]
  \centering
  \includegraphics[width=2.8in]{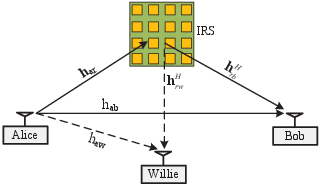}\\
  \caption{IRS-assisted wireless covert communications.}\label{Sys_Sch}
\end{figure}


\section{System Model}

\subsection{Considered Scenario and Assumptions}
As shown in Fig.~\ref{Sys_Sch}, we consider an IRS-assisted covert communication system, where Alice intends to transmit information to Bob covertly with the aid of an IRS, while a warden Willie seeks to detect the existence of this transmission.
We assume that each of Alice, Bob, and Willie is equipped with a single antenna. The IRS is equipped with $N$ passive reflecting elements and the reflection coefficient (including phase shift and reflection amplitude) of each reflecting element can be dynamically adjusted based on the propagation environment.
In addition, it is assumed that the signals reflected by the IRS twice or more are ignored due to the significant path loss\cite{Wu2019Intelligent1}. All channels in the considered system are subject to quasi-static flat-fading.
Specifically, the baseband equivalent channels from Alice to IRS, Bob and Willie are denoted by $\mathbf{h}_{ar}=\sqrt{\chi_{ar}}\mathbf{\bar{h}}_{ar}$, $h_{ab}=\sqrt{\chi_{ab}}\bar{h}_{ab}$, and $h_{aw}=\sqrt{\chi_{aw}}\bar{h}_{aw}$, respectively, while the channels from IRS to Bob and to Willie are denoted by $\mathbf{h}_{rb}^H=\sqrt{\chi_{rb}}\mathbf{\bar{h}}_{rb}^H$ and $\mathbf{h}_{rw}^H= \sqrt{\chi_{rw}}\mathbf{\bar{h}}_{rw}^H$, respectively. In addition, $\chi_{ij}$ denotes the large-scale path loss, where $ij\in\{ar,ab,aw,rb,rw\}$ corresponding to different channels, while $\mathbf{\bar{h}}_{ar}$, $\bar{h}_{ab}$, $\bar{h}_{aw}$, $\mathbf{\bar{h}}_{rb}^H$, and $\mathbf{\bar{h}}_{rw}^H$ are the corresponding small-scale fading coefficients.
We denote $\mathbf{\Theta}=\mathrm{diag}\left(\rho_1e^{j\theta_1},\rho_2e^{j\theta_2},\cdots,\rho_Ne^{j\theta_N}\right)$ as the diagonal reflecting matrix of IRS, where $\theta_n\in[0,2\pi)$ and $\rho_n\in[0,1]$, $n=1,2,\cdots,N$, represent the phase shift and reflection amplitude on the combined incident signal at the $n$-th element, respectively. Furthermore, we assume that the signal transmitted by Alice in the $i$-th channel use is denoted by $x[i]$, $\forall i\in \{1,2,\cdots,L\}$, where $x[i]\thicksim \mathcal{CN}(0,1)$ and $L$ is the total number of channel uses, which is the total number of symbols transmitted over the considered fading block.

\subsection{Binary Hypothesis Testing at Willie}

In this work, we focus on delay-constrained covert communications, i.e., the number of channel uses $L$ is finite. In order to detect the existence of the transmission, Willie is required to distinguish between the following two hypotheses:
\begin{align}\label{Hy_test}
&y_w[i]=
\begin{cases}
n_w[i],&\Hnull,\\
\sqrt{P_a}\left(\mathbf{h}^H_{rw}\mathbf{\Theta}\mathbf{h}_{ar}+h_{aw}\right)x[i]+n_w[i],&\Halt,
\end{cases}
\end{align}
where $y_w[i]$ is the received signal at Willie for the $i$-th channel use, $\Hnull$ denotes the null hypothesis in which Alice does not transmit, and $\Halt$ denotes alternative hypothesis in which Alice transmits information to Bob. In addition, $P_a$ is the transmit power of Alice and $n_w[i]$ is the AWGN at Willie with zero mean and variance $\sigma_w^2$. Following \eqref{Hy_test}, the false alarm rate and miss detection rate at Willie are given by $\mathrm{Pr}\{\Hoalt|\Hnull\}$ and $\mathrm{Pr}\{\Honull|\Halt\}$, respectively, where $\Hoalt$ and $\Honull$ are the binary decisions that infer whether Alice's transmission occurred or not. Then, the total detection error rate at Willie is given by
\begin{align}\label{TDR}
\xi=\pi_0\mathrm{Pr}\{\Hoalt|\Hnull\}+\pi_1\mathrm{Pr}\{\Honull|\Halt\},
\end{align}
where $\pi_0$ and $\pi_1=1-\pi_0$ denote the \emph{priori} probabilities of hypotheses $\Hnull$ and $\Halt$, respectively.
We note that the knowledge of the \emph{priori} probabilities is beneficial to improve Willie's detection performance. In this work, we assume $\pi_0=\pi_1=0.5$ (i.e., equal \emph{priori} probabilities), which has been widely adopted in covert communications (e.g., \cite{Wang2020Secrecy,Zheng2019Multi}).

In covert communications, Willie wishes to minimize its total detection error rate $\xi$ to detect the presence of the transmission. The optimal test that minimizes $\xi$ is the likelihood ratio test, which is given by
\begin{align}\label{like_ratio}
\frac{\mathbb{P}_1\triangleq \prod_{i=1}^L f(y_w[i]|\Halt)}{\mathbb{P}_0\triangleq \prod_{i=1}^L f(y_w[i]|\Hnull)}\mathop{\gtreqless}\limits_{\Honull}^{\Hoalt}1,
\end{align}
where $\mathbb{P}_0$ and $\mathbb{P}_1$ are the likelihood functions of Willie's observation vector over $L$ independent channel uses under $\Hnull$ and $\Halt$, respectively. We have $f(y_w[i]|\Hnull)=\mathcal{CN}(0,\sigma_w^2)$ and $f(y_w[i]|\Halt)\!=\!\mathcal{CN}(0,P_a|\mathbf{h}^H_{rw}\mathbf{\Theta}\mathbf{h}_{ar}\!+\!h_{aw}|^2\!+\!\sigma_w^2)$ as the likelihood function of $y_w[i]$ under $\Hnull$ and $\Halt$, respectively. We note that the optimal detection threshold and the corresponding minimum detection error rate $\xi^*$ at Willie can be derived based on \eqref{like_ratio} \cite{Shihao2018Delay}. However, the resultant expression for $\xi^*$ involves incomplete gamma functions, which is not tractable  for subsequent analysis and design. To overcome this difficulty, we present a lower bound on $\xi^*$, which is given by \cite{Bash2013Limits}
\begin{align}\label{ksi}
\xi^*\geq 1-\sqrt{\frac{1}{2}\mathcal{D}(\mathbb{P}_0|\mathbb{P}_1)},
\end{align}
where $\mathcal{D}(\mathbb{P}_0|\mathbb{P}_1)$ is the KL divergence from $\mathbb{P}_0$ to $\mathbb{P}_1$, and given by \cite{Shihao2018Delay}
\begin{align}\label{D01}
\mathcal{D}(\mathbb{P}_0|\mathbb{P}_1)=&L\Bigg[\ln\left(1+\frac{P_a|\mathbf{h}^H_{rw}\mathbf{\Theta}\mathbf{h}_{ar}+h_{aw}|^2}{\sigma_w^2}\right)\nonumber\\
&~-\frac{P_a|\mathbf{h}^H_{rw}\mathbf{\Theta}\mathbf{h}_{ar}+h_{aw}|^2}{P_a|\mathbf{h}^H_{rw}\mathbf{\Theta}\mathbf{h}_{ar}+h_{aw}|^2+\sigma_w^2}\Bigg].
\end{align}
Following the fact that $\mathbf{h}^H_{rw}\mathbf{\Theta}\mathbf{h}_{ar}=\mathbf{v}^H\mathrm{diag}(\mathbf{h}^H_{rw})\mathbf{h}_{ar}$, $|\mathbf{h}^H_{rw}\mathbf{\Theta}\mathbf{h}_{ar}+h_{aw}|^2$ can be equivalently rewritten as $|\mathbf{v}^H\mathbf{a}\!+\!h_{aw}|^2$, where $\mathbf{v}\!=\![v_1,v_2,\cdots,
v_N]^T$, $\mathbf{a}=\mathrm{diag}(\mathbf{h}^H_{rw})\mathbf{h}_{ar}$ and $v_n=\rho_ne^{-j\theta_n}$, $\forall n$.

In covert communications, $\xi^*\geq 1-\epsilon$ is generally adopted as the covertness constraint, where $\epsilon$ is a small value to determine the required covertness level. As per \eqref{ksi}, we note that $\mathcal{D}(\mathbb{P}_0|\mathbb{P}_1)\leq2\epsilon^2$ is a more stringent constraint than the constraint $\xi^*\geq 1-\epsilon$. As such, in this work we adopt $\mathcal{D}(\mathbb{P}_0|\mathbb{P}_1)\leq2\epsilon^2$ as the required covertness constraint.

\subsection{Transmission from Alice to Bob}

When Alice transmits information, the received signal at Bob for the $i$-th channel use can be expressed as
\begin{align}\label{y_b}
y_b[i]=\sqrt{P_a}\left(\mathbf{h}^H_{rb}\mathbf{\Theta}\mathbf{h}_{ar}+h_{ab}\right)x[i]+n_b[i],
\end{align}
where $n_b[i]$ is the AWGN at Bob with zero mean and variance $\sigma_b^2$. The corresponding SNR at Bob is given by
\begin{align}\label{gammab1}
\gamma_b&=\frac{P_a}{\sigma_b^2}|\mathbf{h}^H_{rb}\mathbf{\Theta}\mathbf{h}_{ar}+h_{ab}|^2=\frac{P_a}{\sigma_b^2}|\mathbf{v}^H\mathbf{b}+h_{ab}|^2,
\end{align}
where $\mathbf{b}=\mathrm{diag}(\mathbf{h}^H_{rb})\mathbf{h}_{ar}$.

We should point out that the decoding error probability $\delta$ at Bob is not negligible for a fixed transmission rate $R$, when the number of channel uses $L$ is finite. As such, the effective throughput, i.e., $LR(1-\delta)$, can be employed to quantify the covert transmission performance of the considered delay-constrained scenario.
We note that the effective throughput increases with $L$. However, Willie would have more observations to detect the covert communication as $L$ increases, which improves his detection performance. We also note that $\mathcal{D}(\mathbb{P}_0|\mathbb{P}_1)$ is a monotonicity increasing function of $P_a|\mathbf{h}^H_{rw}\mathbf{\Theta}\mathbf{h}_{ar}+h_{aw}|^2$.
As such, we can  reduce the value of $P_a|\mathbf{h}^H_{rb}\mathbf{\Theta}\mathbf{h}_{ar}+h_{ab}|^2$ by properly designing the transmit power $P_a$ and IRS reflection beamforming  in order to limit the value of $\mathcal{D}(\mathbb{P}_0|\mathbb{P}_1)$ when $L$ increases, in order to satisfy the covertness constraint.
The recent works \cite{Shihao2018Delay} and \cite{Sun2018Short} revealed that the optimal number of channel uses in the delay-constraint covert communications is the maximum allowable number of channel uses and the effective throughput is an increasing function of $\gamma_b$ for a fixed transmission rate $R$.
As such, in this work we use $\gamma_b$ to evaluate the communication quality from Alice to Bob, and then we aim to maximize $\gamma_b$ subject to the covertness constraint $\mathcal{D}(\mathbb{P}_0|\mathbb{P}_1)\leq2\epsilon^2$ and other practical constraints.

\section{Covert Communication Design with Global  Channel State Information}

In this section, we assume that the CSI of all channels in the considered system is publicly available, which enables us to obtain an upper bound on the performance gain achieved by introducing IRS into covert communications. We note that the acquisition of accurate CSI in IRS-aided communication systems is practically challenging, since the total number of channel coefficients that need to be estimated are significantly increased compared with the case without IRS, while IRS is generally without active transmit RF chains. This challenge was addressed in the literature. For example, the recent works \cite{Liu2020MatrixCal} and \cite{Wei2021ChannelEst} respectively proposed a matrix-calibration based cascaded channel estimation method  and a parallel
factor decomposition based framework, which can accurately estimate the involved channels in IRS-aided systems.


\subsection{Optimization Problem and Perfect Covertness Condition}

Our goal is to maximize the received SNR at Bob by jointly designing the transmit power at Alice and reflect beamforming vector $\mathbf{v}$ (i.e., the phase shifts and reflection amplitudes) at the IRS, subject to the covertness constraint and the maximum transmit power constraint at Alice together with the IRS reflection coefficients constraint. The formulated optimization problem can be written as
\begin{subequations}\label{PF1}
\begin{align}
(\mathrm{P}1):~&\max_{P_a,\mathbf{v}}~P_a|\mathbf{v}^H\mathbf{b}+h_{ab}|^2\label{PF1a}\\
&\mathrm{s.t.}~\mathcal{D}(\mathbb{P}_0|\mathbb{P}_1)\leq 2\epsilon^2,\label{PF1b}\\
&~~~~~P_a\leq P_{\max},\label{PF1c}\\
&~~~~~|v_n|\leq 1,\forall n=1,2,\cdots,N,\label{PF1d}
\end{align}
\end{subequations}
where $\sigma_b^2$ is omitted in the objective function \eqref{PF1a}, since it is a constant term. In addition, \eqref{PF1b} is the covertness constraint and \eqref{PF1d} is due to $\rho_n\in[0,1]$, since the IRS cannot amplify signals.
We note that, following  \cite{Xie2021MaxIRS} and \cite{MursiaRISMA2021}, constraint \eqref{PF1d} implies that the reflection amplitude and phase shift of each IRS element can be independently adjusted over $[0,1]$ and $[0,2\pi)$, respectively.

We should point out that for some specific hardware implementations, IRS reflection amplitudes may depend on its phase shifts, which leads to that the reflection amplitudes and phase shifts are coupled \cite{Abeywickrama2020IRS}. As such, the independent design of IRS reflection amplitudes and phase shifts may suffer from a certain performance degradation caused by this coupling impact. Meanwhile, some existing works (e.g., \cite{Xie2021MaxIRS,MursiaRISMA2021}) expected that near future hardware improvement may support the feasibility that the IRS reflection amplitudes and phase shifts can be adjusted independently.
On the other hand, the formulated optimization problem (P1) jointly designs the reflection amplitudes and phase shifts of the IRS, where only designing the phase shifts of the IRS serves as a special case. In addition, the considered joint design can help us obtain an upper bound on the performance gain achieved by introducing an IRS into covert communications.
Furthermore, it can also help us determine the impact of IRS amplitude control on delay-constraint covert communications.

We note that the considered representative single-antenna setup enables us to identify the fundamental reasons why IRS is beneficial to covert communications and to facilitate us to draw useful insights regarding the impacts of different system parameters on the system performance. The extension of the considered setup to the one with multi-antenna transceivers (e.g., Alice, Bob, and Willie) requires the joint optimization of transmit beamforming and IRS's reflection coefficients. For this joint optimization, the number of data streams transmitted from Alice to Bob and the observation correlation at different antennas should be considered. In addition, the number of antennas at Alice and Willie could significantly affect the optimal design together with the achievable communication covertness of such a multi-antenna system, e.g., the conditions for achieving perfect communication covertness. This is beyond the scope of the paper and such an extension is left for future work.

Before solving the problem (P1), in the following theorem, we first identify the conditions for achieving perfect covertness (i.e., $\mathcal{D}(\mathbb{P}_0|\mathbb{P}_1) =0$) with non-zero transmit power in the considered IRS-assisted covert communication system. We note that when the blocklength is finite, perfect covertness is not achievable in covert communication systems without an IRS (e.g., \cite{Shihao2018Delay}).

\begin{theorem}\label{theorem1}
Perfect covertness can be achieved with non-zero transmit power, i.e., the below optimization problem
\begin{subequations}\label{the1}
\begin{align}
&(\mathrm{P}1')~\max_{P_a,\mathbf{v}}~P_a|\mathbf{v}^H\mathbf{b}+h_{ab}|^2\label{the1a}\\
&\mathrm{s.t.}~P_a|\mathbf{v}^H\mathbf{a}+h_{aw}|^2=0,\label{the1b}\\
&~~~~~P_a\leq P_{\max},\label{the1c}\\
&~~~~~|v_n|\leq 1,\forall n=1,2,\cdots,N.\label{the1d}
\end{align}
\end{subequations}
is feasible, if and only if $\sum_{n=1}^N|a_n|\geq |h_{aw}|$, where $a_n$ is the $n$-th element of $\mathbf{a}=\mathrm{diag}(\mathbf{h}^H_{rw})\mathbf{h}_{ar}$.
\end{theorem}

\begin{IEEEproof}
The detailed proof is provided in Appendix A.
\end{IEEEproof}

We note that $\mathbf{v}^H\mathbf{a}$ is the equivalent channel coefficient of reflect-path from Alice to Willie through the IRS.
We also note that $\sum_{n=1}^N|a_n|=||\mathbf{a}||_1$, which can be equivalently written as $||\mathbf{a}||_1=|\mathbf{v}^H\mathbf{a}|_1$ when $\rho_n=1,\forall n$. In addition, $h_{aw}$ is the channel coefficient of direct-path from Alice to Willie. Thus, Theorem~\ref{theorem1} implies that when the channel quality of the reflect-path is higher than that of the direct-path, perfect covertness can be achieved.
Interestingly, Theorem~\ref{theorem1} also reveals that the additional reflect-path can deteriorate Willie's detection performance (i.e., Willie receives less covert information energy transmitted by Alice), when the reflect beamforming is properly designed. This essentially shows that the IRS is capable of effectively improving the communication covertness, which will be explicitly examined through solving the optimization problem (P1). We should point out that under the condition of achieving perfect covertness as given in Theorem~1, an eavesdropper is not able to wiretap any information from the legitimate transmission, since the eavesdropper cannot receive any useful signal for decoding the confidential information, which indicates that this condition can also be applied to guarantee the physical layer security\cite{Chen2017Survey}.


It should be emphasized that problem $(\mathrm{P}1')$ given in \eqref{the1} is a special case of the problem (P1), where (P1) is non-convex due to the non-concave objective function and the non-convex covertness constraint~\eqref{PF1b}. In the following, we first propose a joint design based on PSCA algorithm to solve (P1) and then we develop a low-complexity solution to balance the computational complexity and the achievable covert communication performance.

\subsection{Joint Transmit Power and Reflect Beamforming Design}



In this subsection, we first transform problem (P1) into a GNCP problem. Then, a PSCA algorithm is developed to solve the resultant problem.
To proceed, we first note that $P_a|\mathbf{v}^H\mathbf{b}+h_{ab}|^2$ in the objective function \eqref{PF1a} and the term $P_a|\mathbf{v}^H\mathbf{a}+h_{aw}|^2$ in the covertness constraint \eqref{PF1b} can be equivalently rewritten as
\begin{align}
P_a\left(\mathbf{v}^H\mathbf{b}\mathbf{b}^H\mathbf{v}+2\mathrm{Re}(\mathbf{v}^H\mathbf{b}h_{ab}^*t)+|h_{ab}|^2|t|^2\right),\label{PCCP_1}\\
P_a\left(\mathbf{v}^H\mathbf{a}\mathbf{a}^H\mathbf{v}+2\mathrm{Re}(\mathbf{v}^H\mathbf{a}h_{aw}^*t)+|h_{aw}|^2|t|^2\right),\label{PCCP_2}
\end{align}
respectively, where the newly introduced slack variable $t$ satisfies $|t|^2=1$. In fact, \eqref{PCCP_1} and \eqref{PCCP_2} can be further rewritten into a quadratic forms $P_a\mathbf{u}^H\mathbf{B}\mathbf{u}$ and $P_a\mathbf{u}^H\mathbf{A}\mathbf{u}$, respectively, where
\begin{align}\label{PCCP_3}
\mathbf{u}\!\!=\!\!\left[\begin{matrix}
   \mathbf{v} \\
   t
\end{matrix}\right],
\mathbf{B}\!\!=\!\!\left[\begin{matrix}
   \mathbf{b}\mathbf{b}^H & \mathbf{b}h_{ab}^*\\
   h_{ab}\mathbf{b}^H & |h_{ab}|^2
\end{matrix}\right],
\mathbf{A}\!\!=\!\!\left[\begin{matrix}
   \mathbf{a}\mathbf{a}^H & \mathbf{a}h_{aw}^*\\
   h_{aw}\mathbf{a}^H & |h_{aw}|^2
\end{matrix}\right].
\end{align}
Following the above transformations, (P1) can be equivalently rewritten as
\begin{subequations}\label{PF1_1}
\begin{align}
&(\mathrm{P}1.1):~\max_{P_a,\mathbf{u}}~P_a\mathbf{u}^H\mathbf{B}\mathbf{u}\label{PF1_1a}\\
&\mathrm{s.t.}~\ln\left(1+\frac{P_a\mathbf{u}^H\mathbf{A}\mathbf{u}}{\sigma_w^2}\right)-\frac{P_a\mathbf{u}^H\mathbf{A}\mathbf{u}}{P_a\mathbf{u}^H\mathbf{A}\mathbf{u}+\sigma_w^2}\leq \frac{2\epsilon^2}{L},\label{PF1_1b}\\
&~~~~~P_a\leq P_{\max},\label{PF1_1c}\\
&~~~~~|u_n|\leq 1,\forall n=1,2,\cdots,N,\label{PF1_1d}\\
&~~~~~|u_{N+1}|= 1,\label{PF1_1e}
\end{align}
\end{subequations}
where \eqref{PF1_1d} is due to $\rho_n\in[0,1]$, while \eqref{PF1_1e} is to guarantee $|t|=1$.
We note that if $\mathbf{u}$ is an optimal solution to the optimization problem (P1.1), $\frac{\mathbf{v}}{t}$ is an optimal solution to the original optimization problem (P1). However, (P1.1) is still difficult to tackle due to the fact that the transmit power variable $P_a$ and the reflect beamforming vector $\mathbf{u}$ are coupled in the objective function \eqref{PF1_1a} and the covertness constraint \eqref{PF1_1b}. Fortunately, $P_a$ is a scale variable and it is constrained by \eqref{PF1_1c}, which allows us to simplify (P1.1) as
\begin{subequations}\label{PF1_2}
\begin{align}
&(\mathrm{P}1.2):~\max_{P_a,\mathbf{w}}~\mathbf{w}^H\mathbf{B}\mathbf{w}\label{PF1_2a}\\
&\mathrm{s.t.}~\ln\left(\frac{\mathbf{w}^H\mathbf{A}\mathbf{w}+\sigma_w^2}{\sigma_w^2}\right)-\frac{\mathbf{w}^H\mathbf{A}\mathbf{w}}{\mathbf{w}^H\mathbf{A}\mathbf{w}+\sigma_w^2}\leq \frac{2\epsilon^2}{L},\label{PF1_2b}\\
&~~~~~P_a\leq P_{\max},\label{PF1_2c}\\
&~~~~~|w_n|\leq \sqrt{P_a},\forall n=1,2,\cdots,N,\label{PF1_2d}\\
&~~~~~|w_{N+1}|= \sqrt{P_a},\label{PF1_2e}
\end{align}
\end{subequations}
where $\mathbf{w}=\sqrt{P_a}\mathbf{u}$.
We note that (P1.2) is NP-hard due to the non-convex constraints \eqref{PF1_2b} and \eqref{PF1_2e}, which generally difficult to tackle directly.
To facilitate the develepment of problem, we first define $\mathbf{W}=\mathbf{w}\mathbf{w}^H$. Then, problem (P1.2) can be recast as
\begin{subequations}\label{PF1_3}
\begin{align}
&(\mathrm{P}1.3):~\max_{P_a,\mathbf{W}}~\mathrm{Tr}(\mathbf{B}\mathbf{W})\label{PF1_3a}\\
&\mathrm{s.t.}~\ln\left(1+\frac{\mathrm{Tr}(\mathbf{A}\mathbf{W})}{\sigma_w^2}\right)-\frac{\mathrm{Tr}(\mathbf{A}\mathbf{W})}{\mathrm{Tr}(\mathbf{A}\mathbf{W})+\sigma_w^2}\leq \frac{2\epsilon^2}{L},\label{PF1_3b}\\
&~~~~~P_a\leq P_{\max},\label{PF1_3c}\\
&~~~~~\mathbf{W}_{n,n}\leq P_a,\forall n=1,2,\cdots,N,\label{PF1_3d}\\
&~~~~~\mathbf{W}_{N+1,N+1}= P_a,\label{PF1_3e}\\
&~~~~~\mathbf{W}\succeq \mathbf{0},\label{PF1_3f}\\
&~~~~~\mathrm{rank}(\mathbf{W})=1,\label{PF1_3g}
\end{align}
\end{subequations}
where \eqref{PF1_3f} and \eqref{PF1_3g} are in (P1.3) to guarantee that $\mathbf{W}=\mathbf{w}\mathbf{w}^H$ holds after the optimization.
We note that the objective function \eqref{PF1_3a} is linear, while \eqref{PF1_3c} and \eqref{PF1_3d} together with \eqref{PF1_3e} are linear constraints. In addition, \eqref{PF1_3f} is a linear matrix inequality (LMI), which are convex with respect to the corresponding optimization variables. In the following, we first present a lemma to determine the convexity of the covertness constraint \eqref{PF1_3b}, and then we tackle the non-convex rank-one constraint \eqref{PF1_3g}.
\begin{lemma}\label{lemma1}
The constraint \eqref{PF1_3b} can be rearranged as
\begin{align}\label{PCCP_4}
&\left(1+\frac{\mathrm{Tr}(\mathbf{A}\mathbf{W})}{\sigma_w^2}\right)\ln\left(1+\frac{\mathrm{Tr}(\mathbf{A}\mathbf{W})}{\sigma_w^2}\right)\nonumber\\
&~~~~~~~~~~-\left(1+\frac{2\epsilon^2}{L}\right)\frac{\mathrm{Tr}(\mathbf{A}\mathbf{W})}{\sigma_w^2}\leq\frac{2\epsilon^2}{L},
\end{align}
which is a convex constraint with respect to $\mathbf{W}$.
\end{lemma}
\begin{IEEEproof}
The detailed proof is provided in Appendix \ref{App_lem1}.
\end{IEEEproof}

In the following, we focus on tackling the non-convex rank-one constraint \eqref{PF1_3g}. In general, problem (P1.3) can be transformed into a convex optimization problem by relaxing the rank-one constraint \eqref{PF1_3g} and the constraint-relaxed problem can be solved by convex optimization tools, such as CVX\cite{Boyd}. However, the optimal $\mathbf{W}$ of the resultant problem may not be rank-one and the corresponding optimal objective value serves as an upper bound on that of problem (P1.3).


In the following, we develop a PSCA iterative algorithm to solve (P1.3), which guarantees that the ultimate convergence solution is a locally optimal solution. To proceed, we first note that the rank-one constraint \eqref{PF1_3g} is equivalent to
\begin{align}\label{PCCP_5}
\mathrm{Tr}(\mathbf{W})-\lambda_{\max}(\mathbf{W})\leq 0,
\end{align}
where $\lambda_{\max}(\mathbf{W})$ is the maximal eigenvalue of $\mathbf{W}$. This follows from the fact that $\mathrm{Tr}(\mathbf{W})-\lambda_{\max}(\mathbf{W})\geq 0$ must hold when $\mathbf{W}\succeq \mathbf{0}$. As such, constraint \eqref{PCCP_5} is equivalent to $\mathrm{Tr}(\mathbf{W})=\lambda_{\max}(\mathbf{W})$, which implies that $\mathbf{W}$ has only one non-zero eigenvalue. It should be emphasized that $\lambda_{\max}(\mathbf{W})$ is a spectral function and is convex with respect to $\mathbf{W}$. Following this fact, the left-hand side (LHS) of the constraint \eqref{PCCP_5} is in the form of a linear function minus a convex function. We note that any convex function is lower-bounded by its first-order approximation at any given point\cite{Boyd}. As a result, the non-convex constraint \eqref{PCCP_5} can be transformed into a more stringent convex constraint for given any feasible solution $\mathbf{\tilde{W}}$.
 Then, the successive convex approximation (SCA) method can be employed to solve the resultant convex optimization problem iteratively.

In applying SCA method, it is difficult to identify an initial feasible $\mathbf{\tilde{W}}$ to the resultant optimization problem, due to the existence of implicit constraint $\mathrm{Tr}(\mathbf{W})-\lambda_{\max}(\mathbf{W})\geq 0$. To overcome this difficulty, we resort to the exact penalty method. Specifically, we first introduce a slack variable $\eta\geq 0$ to enlarge the size of the feasible solution set spanned by constraint \eqref{PCCP_5}. Then we develop a penalty method by adding a slack variable into the objective function. Following the above discussions, we rewrite problem (P1.3) as
\begin{subequations}\label{PF1_4}
\begin{align}
&(\mathrm{P}1.4):~\max_{P_a,\mathbf{W},\eta}~\mathrm{Tr}(\mathbf{B}\mathbf{W})-\tau\eta\label{PF1_4a}\\
&\mathrm{s.t.}~\eqref{PF1_3c},\eqref{PF1_3d},\eqref{PF1_3e},\eqref{PF1_3f},\eqref{PCCP_4},\label{PF1_4b}\\
&~~~~~\mathrm{Tr}(\mathbf{W})-\lambda_{\max}(\mathbf{W})\leq \eta,\label{PF1_4c}\\
&~~~~~\eta\geq 0,\label{PF1_4d}
\end{align}
\end{subequations}
where $\tau>0$ is a penalty parameter. We note that (P1.3) and (P1.4) are equivalent when $\tau>\tau_0$, and thus (P1.4) provides the exact penalty optimization solution to (P1.3).


Now, we turn to address the non-convex constraint \eqref{PF1_4c}.
 Since the spectral function $\lambda_{\max}(\mathbf{W})$ is non-smooth (i.e., not differentiable), we adopt its sub-gradient given by $\mathbf{w}_{\max}\mathbf{w}_{\max}^H$\cite{Hiriart}, where $\mathbf{w}_{\max}$ is the eigenvector associated to the maximum eigenvalue of $\lambda_{\max}(\mathbf{W})$. As such, the first-order restrictive approximation of $\lambda_{\max}(\mathbf{W})$ is replaced by
\begin{align}\label{PCCP_6}
\lambda_{\max}(\mathbf{W})\geq \lambda_{\max}(\mathbf{\tilde{W}})\!+\!\mathrm{Tr}\left(\mathbf{\tilde{w}}_{\max}\mathbf{\tilde{w}}_{\max}^H(\mathbf{W}-\mathbf{\tilde{W}})\right),
\end{align}
where $\mathbf{\tilde{W}}$ is a given feasible point and $\mathbf{\tilde{w}}_{\max}$ is the unit-norm eigenvector corresponding to the
maximum eigenvalue $\lambda_{\max}(\mathbf{\tilde{W}})$ of the matrix $\mathbf{\tilde{W}}$. As such, the constraint \eqref{PF1_4c} can be rewritten as
\begin{align}\label{PCCP_7}
\mathrm{Tr}(\mathbf{W})\!-\!\lambda_{\max}(\mathbf{\tilde{W}})\!-\!\mathrm{Tr}\left(\mathbf{\tilde{w}}_{\max}\mathbf{\tilde{w}}_{\max}^H(\mathbf{W}\!-\!\mathbf{\tilde{W}})\right)\leq \eta.
\end{align}
We should point out that the value of the aforementioned $\tau_0$ can be chosen to be greater than the largest optimal dual variable related to constraint \eqref{PCCP_7} with $\eta=0$ \cite{Lipp2016}.
As per \eqref{PCCP_7}, the optimization problem (P1.4) can be rewritten as
\begin{subequations}\label{PF1_5}
\begin{align}
&(\mathrm{P}1.5):~\max_{P_a,\mathbf{W},\eta}~\mathrm{Tr}(\mathbf{B}\mathbf{W})-\tau\eta\nonumber\\
&\mathrm{s.t.}~\eqref{PF1_3c},\eqref{PF1_3d},\eqref{PF1_3e},\eqref{PF1_3f},\eqref{PCCP_4},\eqref{PF1_4d},\eqref{PCCP_7}.
\end{align}
\end{subequations}

\begin{algorithm}[t]
\caption{PSCA algorithm for Solving (P1.2)}\label{alg1}
\begin{algorithmic}[1]
\STATE Given an initial feasible solution $\mathbf{\tilde{W}}^0$ and an initial penalty parameter $\tau^0$; Given $c>1$ and $\tau_{\max}$; Set $r=0$.
\REPEAT
\STATE {Solve (P1.5) with given a feasible solution $\mathbf{\tilde{W}}^r$ and obtain the current optimal solution $\{\mathbf{W}^{r+1}, P_a^{r+1}, \eta^{r+1}\}$.}
\STATE {Update $\tau^{r+1}=\min\{c\tau^r,\tau_{\max}\}$ and set $\mathbf{\tilde{W}}^{r+1}=\mathbf{W}^{r+1}$; Set the iteration number $r=r+1$.}
\UNTIL {Convergence.}
\end{algorithmic}
\end{algorithm}

We note that the problem (P1.5) is a GNCP problem due to the exponential cone constraint involved in the covertness constraint \eqref{PCCP_4}. For a given penalty parameter $\tau$ and an initial feasible solution $\mathbf{\tilde{W}}$, it can be solved by convex optimization solvers such as CVX\cite{Boyd}. The optimal solution to (P1.5) is also a feasible solution to (P1.4), since the feasible set of (P1.5) is smaller than that of (P1.4). In addition, problem (P1.2) can be tackled by solving (P1.5) iteratively. The detailed iterative algorithm is presented in Algorithm~\ref{alg1}. We note that Algorithm~\ref{alg1} starts with a small value of the penalty parameter $\tau^0$ to put a less emphasis in forcing the rank-one constraint.
Then, the penalty parameter $\tau$ is gradually increased by a constant $c>1$ at each iteration until a large upper bound $\tau_{\max}$ is achieved to guarantee $\eta=0$. We note that Algorithm~\ref{alg1} does not guarantee that the value of objective function always increases with the iteration number, but the objective value will converge. The former is due to the disturbance of the penalty term in the objective function, while the latter is due to the fact that Algorithm~\ref{alg1} is reduced to a standard SCA algorithm when $\tau$ reaches its upper bound $\tau_{\max}$. We note that the convergence of the standard SCA algorithm has been proven in \cite{Li2013Coordinated}.
 {We note that solving a GNCP problem requires a high computational complexity compared to solving other standard convex programs such as
SDP\cite{Tervo2015Optimal}.
 We also note that the complexity of solving a SDP problem with the same size as the problem (P1.5) is $\mathcal{O}\left((N+1)^{6.5})\right)$\cite{Chu2020Intelligent}.
As such, the complexity of the proposed Algorithm~\ref{alg1} is at least on the order of $\mathcal{O}\left(K_1(N+1)^{6.5})\right)$, where $K_1$ is the number of iterations.}

%


\subsection{Low-Complexity Algorithm}
%

In this subsection, we develop a low-complexity two-stage algorithm to strike a balance between the covert transmission performance and computational complexity. Specifically, the IRS's reflection beamforming is designed in the first stage, while Alice's transmit power is determined in the second stage.



\subsubsection{IRS Beamforming Design}
In order to develop a low-complexity IRS beamforming design, as per (P1.1) we note that $\mathcal{D}(\mathbb{P}_0|\mathbb{P}_1)$ in the covertness constraint \eqref{PF1_1b} is a monotonically increasing function of $P_a\mathbf{u}^H\mathbf{A}\mathbf{u}$, which implies that the covertness level is dominated by the received energy at Willie. Following this fact, we adopt the ratio from the received energy at Bob to the received energy at Willie, i.e., $\frac{P_a\mathbf{u}^H\mathbf{B}\mathbf{u}}{P_a\mathbf{u}^H\mathbf{A}\mathbf{u}}$, as our performance metric to design the reflection beamforming, yielding the following optimization problem
\begin{subequations}\label{PF2}
\begin{align}
&(\mathrm{P}2):~\max_{\mathbf{u}}~\frac{\mathbf{u}^H\mathbf{B}\mathbf{u}}{\mathbf{u}^H\mathbf{A}\mathbf{u}}\label{PF2a}\\
&\mathrm{s.t.}~|u_n|\leq 1,\forall n=1,2,\cdots,N,\label{PF2b}\\
&~~~~~|u_{N+1}|= 1.\label{PF2c}
\end{align}
\end{subequations}
We note that the optimization problem (P2) is difficult to tackle directly, since the objective function \eqref{PF2a} is highly non-concave and the unit modules constraint \eqref{PF2c} is non-convex. Furthermore, this problem is quite different from the generalized Rayleigh quotient problem, due to the multiple reflection coefficient constraints that characterize the phase shift and reflection amplitude limits.
In the following, we first derive a lower bound on the objective function \eqref{PF2a}, and then we develop a SCA algorithm to solve (P2) iteratively.
Based on \cite{Guan2020Joint}, a lower bound on \eqref{PF2a} is given by
\begin{align}\label{Low_1}
\frac{\mathbf{u}^H\mathbf{B}\mathbf{u}}{\mathbf{u}^H\mathbf{A}\mathbf{u}}\geq \frac{2\mathrm{Re}(\mathbf{\tilde{u}}^H\mathbf{B}\mathbf{u})}{\mathbf{\tilde{u}}^H\mathbf{A}\mathbf{\tilde{u}}}-\frac{\mathbf{\tilde{u}}^H\mathbf{B}\mathbf{\tilde{u}}}{(\mathbf{\tilde{u}}^H\mathbf{A}\mathbf{\tilde{u}})^2}\mathbf{u}^H\mathbf{A}\mathbf{u},
\end{align}
where $\mathbf{\tilde{u}}$ is a given feasible point.
We note that although the lower bound detailed in \eqref{Low_1} is a concave function of $\mathbf{u}$, it is not conducive to derive a low-complexity analytic expression for $\mathbf{u}$,
As such, we further establish an upper bound on $\mathbf{u}^H\mathbf{A}\mathbf{u}$ \cite{Pan2020Multicell}, as below:
\begin{align}\label{Low_2}
\mathbf{u}^H\mathbf{A}\mathbf{u}\leq \mathbf{u}^H\mathbf{M}\mathbf{u}+2\mathrm{Re}\left(\mathbf{u}^H(\mathbf{A}-\mathbf{M})\mathbf{\tilde{u}}\right)+\mathbf{\tilde{u}}^H\left(\mathbf{M}-\mathbf{A}\right)\mathbf{\tilde{u}},
\end{align}
where $\mathbf{M}=\lambda_{\max}(\mathbf{A})\mathbf{I}_{N+1}$. We recall that $\mathbf{A}$ is a rank-one matrix, it follows that $\lambda_{\max}(\mathbf{A})=\mathbf{\bar{a}}^H\mathbf{\bar{a}}$, where $\mathbf{\bar{a}}=[\mathbf{a}^H ~h_{aw}^*]^H$.
Then, substituting \eqref{Low_2} into \eqref{Low_1} and considering that the values of $||\mathbf{u}||^2$ and $||\mathbf{\tilde{u}}||^2$ are less than or equal to $N+1$, we have
\begin{align}\label{Low_3}
\frac{\mathbf{u}^H\mathbf{B}\mathbf{u}}{\mathbf{u}^H\mathbf{A}\mathbf{u}}\geq 2\mathrm{Re}\left(\mathbf{f}^H\mathbf{u}\right)+\varsigma,
\end{align}
where
\begin{align}
\mathbf{f}&=\left(\frac{\mathbf{B}}{\mathbf{\tilde{u}}^H\mathbf{A}\mathbf{\tilde{u}}}-\frac{\left(\mathbf{A}-\mathbf{\bar{a}}^H\mathbf{\bar{a}}\mathbf{I}_{N+1}\right)\mathbf{\tilde{u}}^H\mathbf{B}\mathbf{\tilde{u}}}{(\mathbf{\tilde{u}}^H\mathbf{A}\mathbf{\tilde{u}})^2}\right)\mathbf{\tilde{u}},\label{Low_4_1}\\
\varsigma&=\frac{\mathbf{\tilde{u}}^H\mathbf{B}\mathbf{\tilde{u}}}{\mathbf{\tilde{u}}^H\mathbf{A}\mathbf{\tilde{u}}}-\frac{2\mathbf{\bar{a}}^H\mathbf{\bar{a}}(N+1)\mathbf{\tilde{u}}^H\mathbf{B}\mathbf{\tilde{u}}}{(\mathbf{\tilde{u}}^H\mathbf{A}\mathbf{\tilde{u}})^2}.\label{varsigama_1}
\end{align}
Then, a lower bound value of optimization problem (P2) can be obtained by solving
\begin{subequations}\label{PF2_1}
\begin{align}
&(\mathrm{P}2.1):~\max_{\mathbf{u}}~2\mathrm{Re}\left(\mathbf{f}^H\mathbf{u}\right)+\varsigma\label{PF2_1a}\\
&\mathrm{s.t.}~|u_n|\leq 1,\forall n=1,2,\cdots,N,\label{PF2_1b}\\
&~~~~~|u_{N+1}|=1.\label{PF2_1c}
\end{align}
\end{subequations}

In order to tackle the optimal solution to the optimization problem (P2.1), we first equivalently rewrite the objective function \eqref{PF2_1a} as
\begin{align}\label{Low_4}
2\mathrm{Re}\left(\sum_{n=1}^{N+1}\left(|f_n||u_n|e^{j(\mathrm{arg}(u_n)-\mathrm{arg}(f_n))}\right)\right)+\varsigma,
\end{align}
where $f_n$ is the $n$-th element of $\mathbf{f}$.
As a result, for a given feasible solution $\mathbf{\tilde{u}}$, the optimal solution to (P2.1) is given by $\mathrm{arg}(u_n)=\mathrm{arg}(f_n)$ and $|u_n|=1$, $\forall n$. It follows that $\mathbf{u}=e^{j\mathrm{arg}(\mathbf{f})}$.
{We note that we always have $|u_n|= 1$, $\forall n$, in this low-complexity design, since otherwise we can always increase the value of $|u_n|$ to further increase the objective function. This also shows the sub-optimality of this design, which will be thoroughly investigated in Section V.}


\subsubsection{Transmit Power Design}

For a given $\mathbf{u}$, the optimization problem (P1.1) is simplified to
\begin{subequations}\label{PF3}
\begin{align}
&(\mathrm{P}3):~\max_{P_a}~P_a\mathbf{u}^H\mathbf{B}\mathbf{u}\label{PF3a}\\
&\mathrm{s.t.}~\ln\left(1+\frac{P_a\mathbf{u}^H\mathbf{A}\mathbf{u}}{\sigma_w^2}\right)-\frac{P_a\mathbf{u}^H\mathbf{A}\mathbf{u}}{P_a\mathbf{u}^H\mathbf{A}\mathbf{u}+\sigma_w^2}\leq \frac{2\epsilon^2}{L},\label{PF3b}\\
&~~~~~P_a\leq P_{\max}.\label{PF3c}
\end{align}
\end{subequations}

We note that $\frac{2\epsilon^2}{L}$ generally is a small value in covert communications. As per $\ln(1+x)\leq x$ for $x>-1$, a conservative approximation of the covertness constraint \eqref{PF3b} is given by
\begin{align}\label{Low_5}
\frac{P_a\mathbf{u}^H\mathbf{A}\mathbf{u}}{\sigma_w^2}-\frac{P_a\mathbf{u}^H\mathbf{A}\mathbf{u}}{P_a\mathbf{u}^H\mathbf{A}\mathbf{u}+\sigma_w^2}\leq \frac{2\epsilon^2}{L}.
\end{align}
Then, the optimization problem (P3) can be recast as
\begin{align}\label{PF3_1}
&(\mathrm{P}3.1):~\max_{P_a}~P_a\mathbf{u}^H\mathbf{B}\mathbf{u}\nonumber\\
&\mathrm{s.t.}~\eqref{PF3b}, ~\eqref{PF3c}, ~\eqref{Low_5}.
\end{align}
One can verify that the LHS of constraint \eqref{Low_5} is a  monotonically increasing function of $P_a$, and the objective function in (P3.1) also monotonically increases with $P_a$. Thus, in the optimal solution to (P3.1) the covertness constraint \eqref{Low_5} holds with equality, which lead to $P_a=\frac{\sigma_w^2\left(\epsilon^2\pm\sqrt{\epsilon^4+2\epsilon^2L}\right)}{L\mathbf{u}^H\mathbf{A}\mathbf{u}}$.
Considering that $\epsilon^2\leq\sqrt{\epsilon^4+2\epsilon^2L}$ and $0\leq P_a\leq P_{\max}$, the optimal transmit power is given by
\begin{align}\label{Low_6}
P_a^\ast=\min\left\{\frac{\sigma_w^2\left(\epsilon^2+\sqrt{\epsilon^4+2\epsilon^2L}\right)}{L\mathbf{u}^H\mathbf{A}\mathbf{u}},P_{\max}\right\}.
\end{align}

We should point out that the adopted approximation in \eqref{Low_5} is to find a low-complexity solution to (P3).
In fact, the optimal $P_a$ to (P3) can be achieved by solving equation \eqref{PF3b} with equality.
However, it is a transcendental equation with respect to $P_a$, which does not facilitate the derivation
of an analytical expression for $P_a$.
We note that the approximation solution shown in \eqref{Low_6} is a high-quality solution to (P3), which is mainly due to the fact that the LHS of constraint \eqref{PF3b} is a monotonicity increasing function of $P_a$ and the value of $\frac{2\epsilon^2}{L}$ is small.
We note that $P_a^*$ in \eqref{Low_6} decreases as $\mathbf{u}^H\mathbf{A}\mathbf{u}$ increases, which verifies the effectiveness of the adopted performance metric in (P2), since it guarantees a relatively small value of $\mathbf{u}^H\mathbf{A}\mathbf{u}$.
We also note that $P_a^*$ decreases as the required covertness level $\epsilon$ decreases, which is consistent with our intuition.

The proposed low-complexity two-stage algorithm is summarized in Algorithm~\ref{alg2}. We note that in the first stage we design the reflection beamforming without considering the covertness constraint, i.e., the covertness constraint is not involved in the optimization problem (P2). In the second stage, Alice's transmit power is determined to explicitly ensure the covertness constraint. {We note that the main computational complexity of Algorithm~\ref{alg2} comes from calculating $\mathbf{f}$ in step $3$ and $P_a$ in step $6$. We observe from \eqref{Low_2} and \eqref{Low_6} that the complexity of calculating $\mathbf{f}$ and $P_a$ mainly depends on the calculation of the quadratic form, which is on the order of $\mathcal{O}\left((N+1)^2\right)$. As such, the total computational complexity of Algorithm~\ref{alg2} is given by $\mathcal{O}\left(K_2(N+1)^2\right)$, where $K_2$ is the number of iterations at the first stage.
We note that the complexity of the proposed Algorithm~\ref{alg2} is much lower than that of the proposed Algorithm~\ref{alg1}.}

We should point out that the aforementioned low-complexity two-stage algorithm cannot obtain the perfect covertness with non-zero transmit power detailed in the optimization problem $\mathrm{P}1'$. This is due to the fact that we have performed multiple lower bound approximation operations on the objective function \eqref{PF2a} and Alice's transmit power approaches to $0$ as $\epsilon \rightarrow 0$.
As such, in the following we develop a low-complexity solution to $(\mathrm{P}1')$. To this end, we first recall that, as per Theorem~\ref{theorem1}, the perfect covertness can be achieved when  $\sum_{n=1}^N|a_n|\geq|h_{aw}|$. Following this fact, we can equivalently rewrite $\sum_{n=1}^N|a_n|\geq |h_{aw}|$ as $\sum_{n=1}^N|a_n|= \kappa|h_{aw}|$, where $\kappa\geq 1$ is a scale factor. Then, we can see that $\{v_n=e^{j(\arg(a_n)-\pi-\arg(h_{aw}))}, \forall n, P_a=P_{\max}\}$ is the optimal solution to $(\mathrm{P}1')$ when $\kappa=1$ and  $\{v_n=\frac{1}{\kappa}e^{j(\arg(a_n)-\pi-\arg(h_{aw}))}, \forall n, P_a=P_{\max}\}$ is a feasible solution to $(\mathrm{P}1')$ for $\kappa>1$.

\begin{algorithm}[t]
\caption{Proposed Low-Complexity Algorithm}\label{alg2}
\begin{algorithmic}[1]
\STATE Given an initial feasible solution $\mathbf{\tilde{u}}^0$ and set iteration index $r=0$.
\REPEAT
\STATE {Compute $\mathbf{u}=e^{j\mathrm{arg}(\mathbf{f})}$ to obtain the current optimal solution of problem (P2.1).}
\STATE {Update $\mathbf{\tilde{u}}^r=\mathbf{u}$ and set $r=r+1$.}
\UNTIL {Convergence.}
\STATE {Compute $P_a$ according to \eqref{Low_6}.}
\end{algorithmic}
\end{algorithm}


\section{Covert Communication Design without Willie's instantaneous CSI}

If Willie is not a legitimate user in the considered system for other service, it may be difficult to obtain his instantaneous CSI. As such, in this section we consider that Alice and the IRS only
know that $h_{aw} \sim \mathcal{CN}(0,\chi_{aw})$ and $h_{rw_n} \sim \mathcal{CN}(0,\chi_{rw})$, where $h_{rw_n}$ is the $n$-th element of $\mathbf{h}_{rw}$, but they do not know the instantaneous realizations of $h_{aw}$ or $\mathbf{h}_{rw}$. From a conservative point of view, we assume that Willie knows the instantaneous $h_{aw}$ and $\mathbf{h}_{rw}$.

{\subsection{Expression for Covertness Constraint}}
As per \eqref{D01}, the covertness constraint $\mathcal{D}(\mathbb{P}_0|\mathbb{P}_1)\leq2\epsilon^2$ depends on $h_{aw}$ and $\mathbf{h}_{rw}$. As such, we consider the expected value of $\mathcal{D}(\mathbb{P}_0|\mathbb{P}_1)$ over all realizations of $h_{aw}$ and $\mathbf{h}_{rw}$ as the measure of covertness. Then, the covertness constraint can be rewritten as $\mathbb{E}_X\left[\mathcal{D}(\mathbb{P}_0|\mathbb{P}_1)\right]\leq2\epsilon^2$, where
\begin{align}\label{Sp2_1}
&\mathbb{E}_X\left[\mathcal{D}(\mathbb{P}_0|\mathbb{P}_1)\right]=\mathbb{E}_X\Bigg\{L\Bigg[\ln\left(1+\frac{P_aX}{\sigma_w^2}\right)-\frac{P_aX}{P_aX+\sigma_w^2}\Bigg]\Bigg\},
\end{align}
and $X\triangleq|\mathbf{h}^H_{rw}\mathbf{\Theta}\mathbf{h}_{ar}+h_{aw}|^2$. {We note that the use of $\mathbb{E}_X\left[\mathcal{D}(\mathbb{P}_0|\mathbb{P}_1)\right]\leq2\epsilon^2$ as the covertness constraint lies in the fact that Willie knows the instantaneous $h_{aw}$ and $\mathbf{h}_{rw}$. This fact enables Willie to vary his detection threshold for each instantaneous realization of $h_{aw}$ or $\mathbf{h}_{rw}$, such that $\mathbb{E}_X\left[\mathcal{D}(\mathbb{P}_0|\mathbb{P}_1)\right]$ can be used to provide a lower bound on Willie's minimum detection error rate $\xi^\ast$. We note that, if Willie uses a fixed detection threshold for all the realizations of $h_{aw}$ and $\mathbf{h}_{rw}$, $\mathbb{E}_X\left[\mathcal{D}(\mathbb{P}_0|\mathbb{P}_1)\right]\leq2\epsilon^2$ should not be used as the covertness constraint.} Since the expression of $\mathbb{E}_X\left[\mathcal{D}(\mathbb{P}_0|\mathbb{P}_1)\right]$ is very complex, an exact analytical expression of $\mathbb{E}_X\left[\mathcal{D}(\mathbb{P}_0|\mathbb{P}_1)\right]$ is difficult to obtained directly.
In the following, we present a theorem to determine the analytical expression of $\mathbb{E}_X\left[\mathcal{D}(\mathbb{P}_0|\mathbb{P}_1)\right]\leq2\epsilon^2$.
\begin{theorem}\label{theorem2}
The covertness constraint $\mathbb{E}_X\left[\mathcal{D}(\mathbb{P}_0|\mathbb{P}_1)\right]\leq2\epsilon^2$ can be equivalently rewritten as
\begin{align}\label{Sp2_6}
\frac{P_a}{\sigma_w^2}\left(\chi_{rw}\sum_{n=1}^N\rho_n^2|h_{ar_n}|^2+\chi_{aw}\right)\leq \bar{\epsilon},
\end{align}
where $\bar{\epsilon}$ is the solution to
\begin{align}\label{Sp2_6_2}
\left(1+\frac{1}{\bar{\epsilon}}\right)e^{\frac{1}{\bar{\epsilon}}}E_1\left(\frac{1}{\bar{\epsilon}}\right)-1- \frac{2\epsilon^2}{L}=0,
\end{align}
which can be obtained via the bisection method\cite{Boyd}, where $E_1(x)=\int_x^{\infty}\frac{e^{-t}}{t}dt$ is an exponential integral function.
\end{theorem}
\begin{IEEEproof}
The detailed proof is provided in Appendix~\ref{App_the2}.
\end{IEEEproof}

We note that  $\mathbb{E}_X\left[\mathcal{D}(\mathbb{P}_0|\mathbb{P}_1)\right]\leq2\epsilon^2$ is a new covertness constraint in the context of covert communications when Willie's instantaneous CSI is not available. Theorem~\ref{theorem2} equivalently transforms the mathematically intractable covertness constraint $\mathbb{E}_X\left[\mathcal{D}(\mathbb{P}_0|\mathbb{P}_1)\right]\leq2\epsilon^2$ into the constraint \eqref{Sp2_6}, which facilitates us determine the optimal IRS's reflection beamforming and Alice's transmit power in the next subsection.

\subsection{Optimal Design without Willie's Instantaneous CSI}
Following Theorem~\ref{theorem2}, when Willie's instantaneous CSI is not available, the optimal design of the IRS's reflection beamforming and Alice's transmit power is formulated as
\begin{subequations}\label{PF5}
\begin{align}
&(\mathrm{P}5):~\max_{P_a,\bm{\rho},\bm{\theta}}~P_a\left|\sum_{n=1}^N\rho_n|b_n| e^{j\left(\theta_n+\arg(b_n)\right)}+h_{ab}\right|^2\label{PF5a}\\
&\mathrm{s.t.}~P_a\left(\chi_{rw}\sum_{n=1}^N\rho_n^2|h_{ar_n}|^2+\chi_{aw}\right)\leq \bar{\epsilon}\sigma_w^2,\label{PF5b}\\
&~~~~~P_a\leq P_{\max},\label{PF5c}\\
&~~~~~0\leq\rho_n\leq 1,\forall n=1,2,\cdots,N,\label{PF5d}\\
&~~~~~0\leq\theta_n< 2\pi, \forall n=1,2,\cdots,N,\label{PF5e}
\end{align}
\end{subequations}
where $b_n$ is the $n$-th element of $\mathbf{b}$, $\bm{\rho}=[\rho_1,\rho_2,\cdots,\rho_N]^T$ and $\bm{\theta}=[\theta_1,\theta_2,\cdots,\theta_N]^T$.
Interestingly, we find that covertness constraint \eqref{PF5b} is independent of the phase shift $\bm{\theta}$, which implies that the optimal phase shift strategy at the IRS is to maximize Bob's SNR when IRS does not know Willie's instantaneous CSI. Following the fact that
\begin{align}\label{Sp2_7_1}
&\left|\sum_{n=1}^N\rho_n|b_n| e^{j\left(\theta_n+\arg(b_n)\right)}+h_{ab}\right|\nonumber\\
&~~~~\leq \sum_{n=1}^N\left|\rho_n|b_n| e^{j\left(\theta_n+\arg(b_n)\right)}\right|+|h_{ab}e^{j\arg(h_{ab})}|,
\end{align}
where the equality holds when $\theta_n+\arg(b_n)=\arg(h_{ab}),\forall n$,
we can conclude that the optimal $\bm{\theta}$ to the optimization problem in (P5) is given by
\begin{align}\label{Sp2_7}
\theta_n^*=\arg(h_{ab})-\arg(b_n), \forall n.
\end{align}
Substituting \eqref{Sp2_7} into (P5) and performing the root operation on the objective function \eqref{PF5a}, we have
\vspace{-0.2cm}
\begin{align}\label{PF5_1}
&(\mathrm{P}5.1):~\max_{P_a,\bm{\rho}}~\sqrt{P_a}\left(\sum_{n=1}^N\rho_n|b_n|+|h_{ab}|\right)\nonumber\\
&\mathrm{s.t.}~\eqref{PF5b},\eqref{PF5c},\eqref{PF5d}.
\end{align}

In the following, we present the optimal solution to the optimization problem (P5.1) by considering three possible types of the amplitude regulator to be adapted in the IRS, i.e., different constraints on the IRS's amplitude coefficients in three practical scenarios.

\setcounter{subsubsection}{0}
\subsubsection{$\rho_n=1,\forall n$} In this scenarios, $\bm{\rho}$ is fixed to be $1$, which means that only the phase shifts in the IRS can be controlled. Then, in the problem (P5.1), only Alice's transmit power need to be determined, of which the optimal value is given by
\begin{align}\label{Sp2_8}
P_a^\ast=\min\left\{\frac{\bar{\epsilon}\sigma_w^2}{\chi_{rw}||\mathbf{h}_{ar}||^2+\chi_{aw}},P_{\max}\right\}.
\end{align}

\subsubsection{$\rho_n=\rho_0,\forall n$}
In this scenario, all the elements of the IRS share a common amplitude coefficient controller due to hardware limitations. As such, we need to jointly design $\rho_0$ and $P_a$ in the optimization problem (P5.1). To this end, we first rewrite (P5.1) as
\begin{subequations}\label{PF5_2}
\begin{align}
&(\mathrm{P}5.2):~\max_{P_a,\rho_0}~\sqrt{P_a}\left(\rho_0||\mathbf{b}||_1+|h_{ab}|\right)\label{PF5_2a}\\
&\mathrm{s.t.}~P_a\left(\chi_{rw}\rho_0^2||\mathbf{h}_{ar}||^2+\chi_{aw}\right)\leq \bar{\epsilon}\sigma_w^2,\label{PF5_2b}\\
&~~~~~P_a\leq P_{\max},\label{PF5_2c}\\
&~~~~~0\leq\rho_0\leq 1.\label{PF5_2d}
\end{align}
\end{subequations}
We note that constraint \eqref{PF5_2b} must hold with equality at the optimal solution, otherwise we can always increase $P_a$ or $\rho_0$ to further improve the objective function. As such, the optimal value of the transmit power $P_a$ can be expressed as
\begin{align}\label{Sp2_9}
&P_a^\ast=
\begin{cases}
P_{\max},&\frac{\bar{\epsilon}\sigma_w^2}{\left(\chi_{rw}\rho_0^2||\mathbf{h}_{ar}||^2+\chi_{aw}\right)}>P_{\max}
,\\
\frac{\bar{\epsilon}\sigma_w^2}{\left(\chi_{rw}\rho_0^2||\mathbf{h}_{ar}||^2+\chi_{aw}\right)},&\frac{\bar{\epsilon}\sigma_w^2}{\left(\chi_{rw}\rho_0^2||\mathbf{h}_{ar}||^2+\chi_{aw}\right)}\leq P_{\max}.
\end{cases}
\end{align}

We should point out that increasing the values of $P_a$ and $\rho_0$ can improve the communication quality from Alice to Bob, but may decrease the achievable communication covertness. In addition, we observe from \eqref{Sp2_9} that $P_a$ decreases with $\rho_0$ when $\frac{\bar{\epsilon}\sigma_w^2}{\left(\chi_{rw}\rho_0^2||\mathbf{h}_{ar}||^2+\chi_{aw}\right)}\leq P_{\max}$ holds. As such, the transmit power $P_a$ and the IRS amplitude coefficient $\rho_0$ should be carefully designed to balance the communication quality and communication covertness.
As per \eqref{Sp2_9}, we consider the following two cases.
\paragraph{$P_a^\ast=P_{\max}$} In this case, (P5.2) can be recast as
\begin{subequations}\label{PF5_3}
\begin{align}
&(\mathrm{P}5.3):~\max_{\rho_0}~\sqrt{P_{\max}}\left(\rho_0||\mathbf{b}||_1+|h_{ab}|\right)\label{PF5_3a}\\
&\mathrm{s.t.}~0\leq\rho_0\leq 1,~\rho_0^2\leq \frac{\frac{\bar{\epsilon}\sigma_w^2}{P_{\max}}-\chi_{aw}}{\chi_{rw}||\mathbf{h}_{ar}||^2}.\label{PF5_3b}
\end{align}
\end{subequations}
Since the objective function \eqref{PF5_3a} is a linear function of $\rho_0$, the optimal $\rho_0$ must be on the boundary of \eqref{PF5_3b}. Thus, the optimal $\rho_0$ is given by
\begin{align}\label{Sp2_10}
\rho_0^\ast=\min\left\{1,\sqrt{\frac{\frac{\bar{\epsilon}\sigma_w^2}{P_{\max}}-\chi_{aw}}{\chi_{rw}||\mathbf{h}_{ar}||^2}}\right\}.
\end{align}

We observe from \eqref{Sp2_10} that the optimal $\rho_0$ decreases as $\chi_{aw}$ or $\chi_{rw}$ increases, since as the channel quality from Alice or IRS to Willie is improved, it becomes easier for Willie to make correct decisions on the detection of Alice's transmission. As such, IRS will reduce its amplitudes to maintain the same level of covertness.

\paragraph{$P_a^*=\frac{\bar{\epsilon}\sigma_w^2}{\chi_{rw}\rho_0^2||\mathbf{h}_{ar}||^2+\chi_{aw}}$} In this case, (P5.2) can be rewritten as
\begin{subequations}\label{PF5_4}
\begin{align}
&(\mathrm{P}5.4):~\max_{\rho_0}~\frac{\sqrt{\bar{\epsilon}\sigma_w^2}\left(\rho_0||\mathbf{b}||_1+|h_{ab}|\right)}{\sqrt{\chi_{rw}\rho_0^2||\mathbf{h}_{ar}||^2+\chi_{aw}}}
\label{PF5_4a}\\
&\mathrm{s.t.}~0\leq\rho_0\leq 1,\rho_0^2\geq \frac{\frac{\bar{\epsilon}\sigma_w^2}{P_{\max}}-\chi_{aw}}{\chi_{rw}||\mathbf{h}_{ar}||^2}.\label{PF5_4b}
\end{align}
\end{subequations}
Since (P5.4) is a univariate optimization problem, its optimal solution must be attained either at the stationary point of the objective function or on the boundary of the feasible set, where the stationary point of \eqref{PF5_4a} is given by
\vspace{-0.2cm}
\begin{align}\label{Sp2_11}
\rho_0^s=\frac{\chi_{aw}||\mathbf{b}||_1}{\chi_{rw}|h_{ab}|||\mathbf{h}_{ar}||^2}.
\end{align}
Thus, the optimal $\rho_0$ to problem (P5.4) can be obtained by checking the objective values of the feasible stationary point and endpoints of the constraint \eqref{PF5_4b}.

Finally, to determine the solution to (P5.2) in this scenario, we compare the achieved objective function values in the aforementioned two cases and choose the candidate optimal solution with the higher objective function value as the optimal $\rho_0$. Then, the optimal $P_a$ can be obtained as per \eqref{Sp2_9}.

%



\subsubsection{$0\leq\rho_n\leq 1,\forall n$} In this scenario, we have the general case of jointly optimizing $P_a$ and $\bm{\rho}$. We note that the optimization problem (P5.1) is non-convex due to the non-concave objective function and the non-convex constraint \eqref{PF5b}. Fortunately, it can be equivalently transformed into the following convex form:
\begin{subequations}\label{PF5_5}
\begin{align}
&(\mathrm{P}5.5):~\max_{P_a,\bm{\bar{\rho}}}~\bm{\bar{\rho}}^T\mathbf{\bar{b}}+\sqrt{P_a}|h_{ab}|\label{PF5_5a}\\
&\mathrm{s.t.}~\chi_{rw}\bm{\bar{\rho}}^T\mathbf{H}_{ar}\bm{\bar{\rho}}+P_a\chi_{aw}\leq \bar{\epsilon}\sigma_w^2,\label{PF5_5b}\\
&~~~~~P_a\leq P_{\max},\label{PF5_5c}\\
&~~~~~0\leq\bar{\rho}_n\leq \sqrt{P_a},\forall n=1,2,\cdots,N.\label{PF5_5d}
\end{align}
\end{subequations}
where $\bm{\bar{\rho}}=\sqrt{P_a}\bm{\rho}$, $\mathbf{\bar{b}}=[|b_1|,|b_2|,\cdots,|b_N|]^T$, $\bar{\rho}_n$ is the $n$-th element of $\bm{\bar{\rho}}$, and $\mathbf{H}_{ar}\!=\!\mathrm{diag}\big(|h_{ar_1}|^2,|h_{ar_2}|^2,\cdots,|h_{ar_N}|^2\big)$.
We note that (P5.5) is a convex optimization problem,
which can be efficiently solved by the current convex optimization solver CVX\cite{Boyd}. Then, the optimal $\bm{\rho}$ to the  problem (P5.1) can be recovered by $\frac{\bm{\bar{\rho}}}{\sqrt{P_a}}$.

\section{Numerical Results}

In this section, we provide numerical results to evaluate the performance of the IRS-assisted covert communication system with our proposed designs.
We consider a three-dimensional coordinate system, Alice, IRS, Bob, and Willie are respectively located at $(0, 5,5)$ meter $(\mathrm{m})$, $(100,0,5)~\mathrm{m}$, $(70,10,0)~\mathrm{m}$, and $(100,10,0)~\mathrm{m}$.
In addition, we assume that IRS is equipped with a uniform rectangular array (URA) with $N=N_xN_z$, where $N_x$ and $N_z$ are the number of reflecting elements along the $x$-axis and $z$-axis, respectively.
{Note that all results presented below are averaged over $1000$ independent channel realizations. Considering that IRS is deployed generally with the knowledge of Alice's location, the channel realizations from Alice to IRS are randomly drawn from Rician fading with a Rician factor of $5~\mathrm{dB}$, while all other channel realizations are drawn from Rayleigh fading.}
The large-scale path loss from node $i$ to node $j$ is denoted as $\chi_{ij}=\beta_0\left(\frac{d_{ij}}{d_0}\right)^{-\alpha_{ij}}$, where $\beta_0$ is the channel power gain at the reference distance $1~\mathrm{m}$, $d_{ij}$ is the distance between node $i$ and node $j$, and $\alpha_{ij}$ is the corresponding path loss exponent. Specifically, the path loss exponents are set as $\alpha_{ar}=2.4$, $\alpha_{ab}=4.2$, $\alpha_{aw}=4.2$, $\alpha_{rb}=3$, and $\alpha_{rw}=3$.
Unless stated otherwise, the remaining system parameters are set as follows: $\beta_0=-30~\mathrm{dB}$, $P_{\max}=36~\mathrm{dBm}$, $L=100$, $\sigma_b^2=\sigma_w^2=-80~\mathrm{dBm}$, and $N_x=5$.



\subsection{With Global CSI}

In this subsection, we present numerical results to evaluate the Bob's SNR achieved by our proposed PSCA algorithm and low-complexity algorithm. The proposed design algorithms and the corresponding benchmark schemes are denoted as follows.
\begin{itemize}
\item Upper bound: Achieved by solving problem (P1.3) by relaxing the rank-one constraint.

\item PSCA algorithm ($\rho_n\leq 1, \forall n$):
Jointly design Alice's transmit power as well as phase shift and reflection amplitude of each IRS reflecting element.

\item PSCA algorithm ($\rho_n= 1, \forall n$): 
Jointly design Alice's transmit power and phase shift of each IRS reflection element, where IRS reflection amplitudes are fixed to $1$.

\item Low-complexity algorithm: Proposed Algorithm~\ref{alg2}.

\item Without IRS: No IRS in the considered system and only Alice's transmit power $P_a$ is designed.
\end{itemize}

\begin{figure}[!t]
  \centering
  \includegraphics[width=3.4in, height=2.6in]{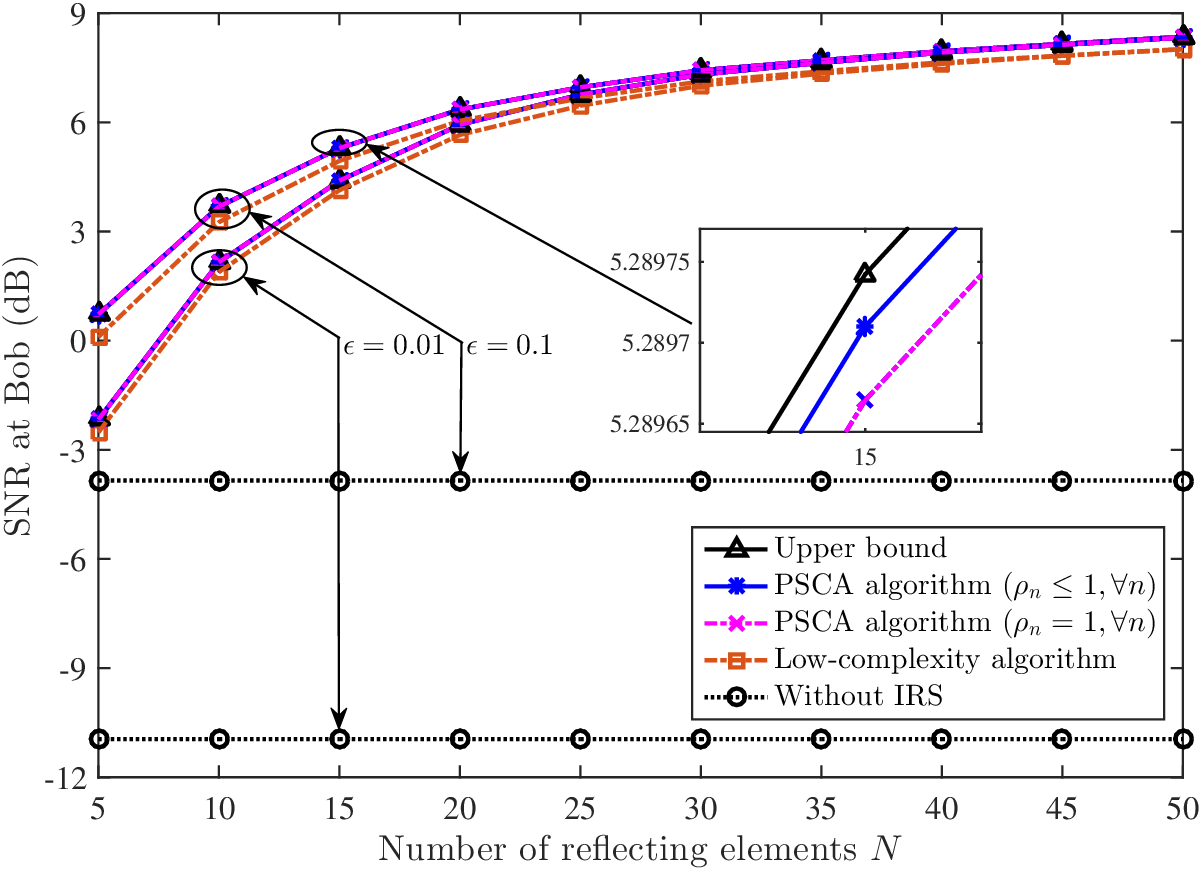}
  \caption{Bob's SNR versus the number of reflecting elements at IRS for different values of the covertness level $\epsilon$.}\label{Num_SNR}
\end{figure}
\begin{figure}[!t]
  \centering
  \includegraphics[width=3.4in, height=2.6in]{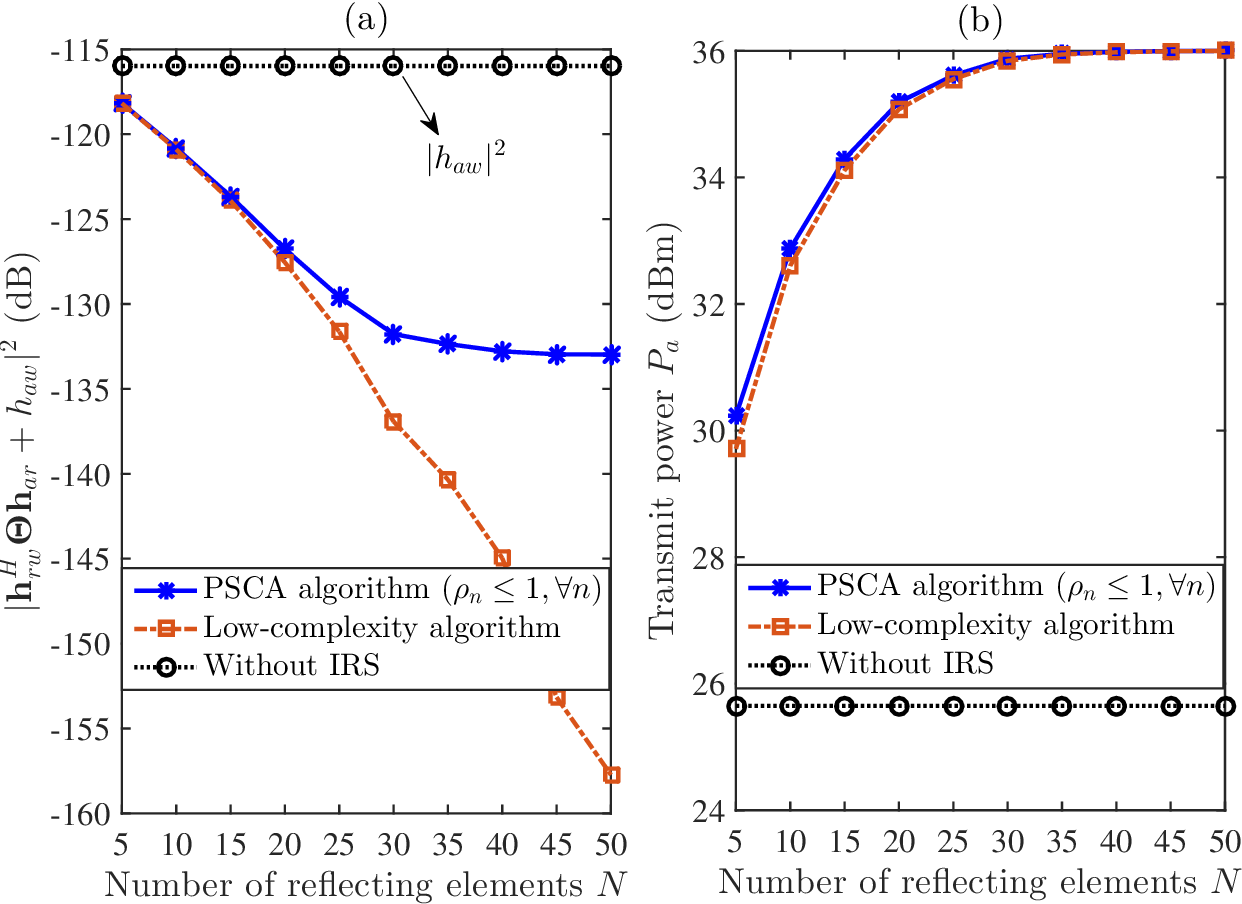}
  \caption{The value of $|\mathbf{h}^H_{rw}\mathbf{\Theta}\mathbf{h}_{ar}+h_{aw}|^2$ at Willie versus the number of reflecting elements at IRS for covertness level $\epsilon=0.1$.}\label{Energy_Willie}
\end{figure}

In Fig.~\ref{Num_SNR}, we plot the received SNR at Bob achieved by different algorithms versus the number of IRS elements for different values of the covertness level $\epsilon$. In this figure, we first observe that Bob's SNR decreases as $\epsilon$ decreases for all schemes. This is due to the fact that the covertness constraint becomes more stringent as $\epsilon$ decreases. As expected, the developed algorithms with IRS significantly outperform the algorithm without IRS in terms of achieving a higher SNR at Bob, which demonstrates the benefits of introducing IRS into covert communications. We also observe that Bob's SNR achieved by all the algorithms with IRS increases with the number of reflecting elements $N$. As a larger $N$ not only enables Bob to receive a stronger reflection signal from the IRS, but also makes the covertness constraint easier to be satisfied, which will be confirmed in Fig.~\ref{Energy_Willie}. In addition, Bob's SNR achieved by PSCA algorithm ($0\leq \rho_n\leq1, \forall n$) and PSCA algorithm ($\rho_n=1, \forall n$) approaches that of the upper bound, which demonstrates that the proposed PSCA algorithms can achieve near-optimal performance. This also implies that the performance gain achieved by adjusting IRS's phase shifts is superior to that achieved by varying IRS's reflection amplitudes in the context of IRS-assisted covert communications with perfect CSI.
This is because the IRS reflected channel suffers form the effect of double path loss \cite{Wu2021IRStutorial}. As such, adjusting the IRS reflection amplitudes would significantly reduce the SNR at Bob. In addition, considering that the global CSI is available, IRS can adjust its phase shifts to ensure that the privacy information power leaked to Willie is relatively small, while guaranteeing a certain communication quality. Following these facts, IRS prefers to adjust its phase shifts rather than its amplitudes for the case with global CSI available.
In this figure, it can be observed that Bob's SNR obtained by the low-complexity algorithm is slightly lower than that obtained by the PSCA algorithms, which shows that the proposed low-complexity algorithm can effectively strike a good balance between the covert communication performance and computational complexity.

In Fig.~\ref{Energy_Willie}, we plot the value of $|\mathbf{h}^H_{rw}\mathbf{\Theta}\mathbf{h}_{ar}+h_{aw}|^2$, which is the channel power gain received at Willie, to reveal the fundamental reason why IRS can improve the covert communication performance. Recall that we adopt $\mathcal{D}(\mathbb{P}_0|\mathbb{P}_1)\leq 2\epsilon^2$ as the covertness constraint, where $\mathcal{D}(\mathbb{P}_0|\mathbb{P}_1)$ is an increasing function of $P_a|\mathbf{h}^H_{rw}\mathbf{\Theta}\mathbf{h}_{ar}+h_{aw}|^2$ (i.e., the received energy at Willie), which implies that the covertness constraint is dominated by $P_a|\mathbf{h}^H_{rw}\mathbf{\Theta}\mathbf{h}_{ar}+h_{aw}|^2$. In Fig.~\ref{Energy_Willie}(a), we first observe that $|\mathbf{h}^H_{rw}\mathbf{\Theta}\mathbf{h}_{ar}+h_{aw}|^2$ achieved by the algorithms with IRS is always lower than that of the algorithm without IRS,
which indicates that with the aid of an IRS, Alice can use a higher transmit power to improve the covert communication performance while meeting the same covertness constraint. This also shows that the IRS not only enhances the transmission from Alice to Bob, but also deteriorates Willie's detection performance. In Fig.~\ref{Energy_Willie}(a), we also observe that the value of $|\mathbf{h}^H_{rw}\mathbf{\Theta}\mathbf{h}_{ar}+h_{aw}|^2$ achieved by the PSCA algorithms is larger than that achieved by the low-complexity algorithm. This is
the main reason why the PSCA algorithms can slightly outperform the low-complexity algorithm.

\begin{figure}[!t]
  \centering
  \includegraphics[width=3.4in, height=2.6in]{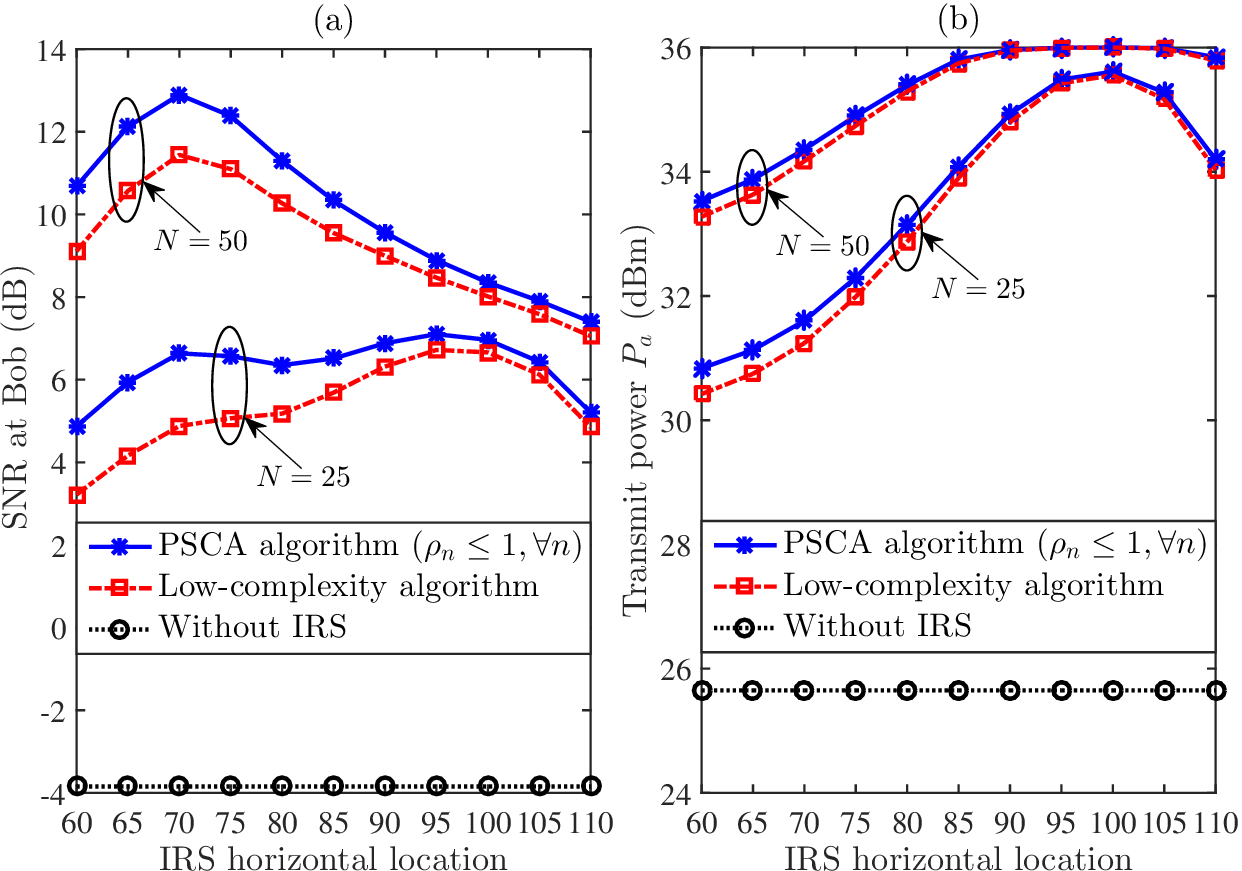}
  \caption{Received SNR at Bob and transmit power at Alice versus IRS horizontal location for covertness level $\epsilon=0.1$.}\label{Location_IRS}
\end{figure}

In Fig.~\ref{Location_IRS}, we plot Bob's SNR and Alice's transmit power $P_a$ achieved by different schemes versus the IRS horizontal location for different values of the reflecting element number $N$. In Fig.~\ref{Location_IRS}(a), we first observe that with a larger $N$, e.g., $N=50$, the optimal horizontal location of the IRS will be closer to Bob, while the optimal horizontal location of the IRS moves closer to Willie as $N$ decreases. The is because the covertness constraint becomes easier to be satisfied as $N$ increases. In general, this figure shows that the optimal horizontal location of the IRS is between Alice and Bob to strike a tradeoff between the communication quality from Alice to Bob and the covertness constraint.
Surprisingly, in Fig.~\ref{Location_IRS}(b) we observe that the transmit power $P_a$ gradually increases as the IRS moves closer to Willie, which further verifies that IRS can reduce Willie's channel power gain $|\mathbf{h}^H_{rw}\mathbf{\Theta}\mathbf{h}_{ar}+h_{aw}|^2$ and thus degrade its detection performance, especially when the IRS is close to Willie.

\begin{figure}[!t]
  \centering
  \includegraphics[width=3.4in, height=2.6in]{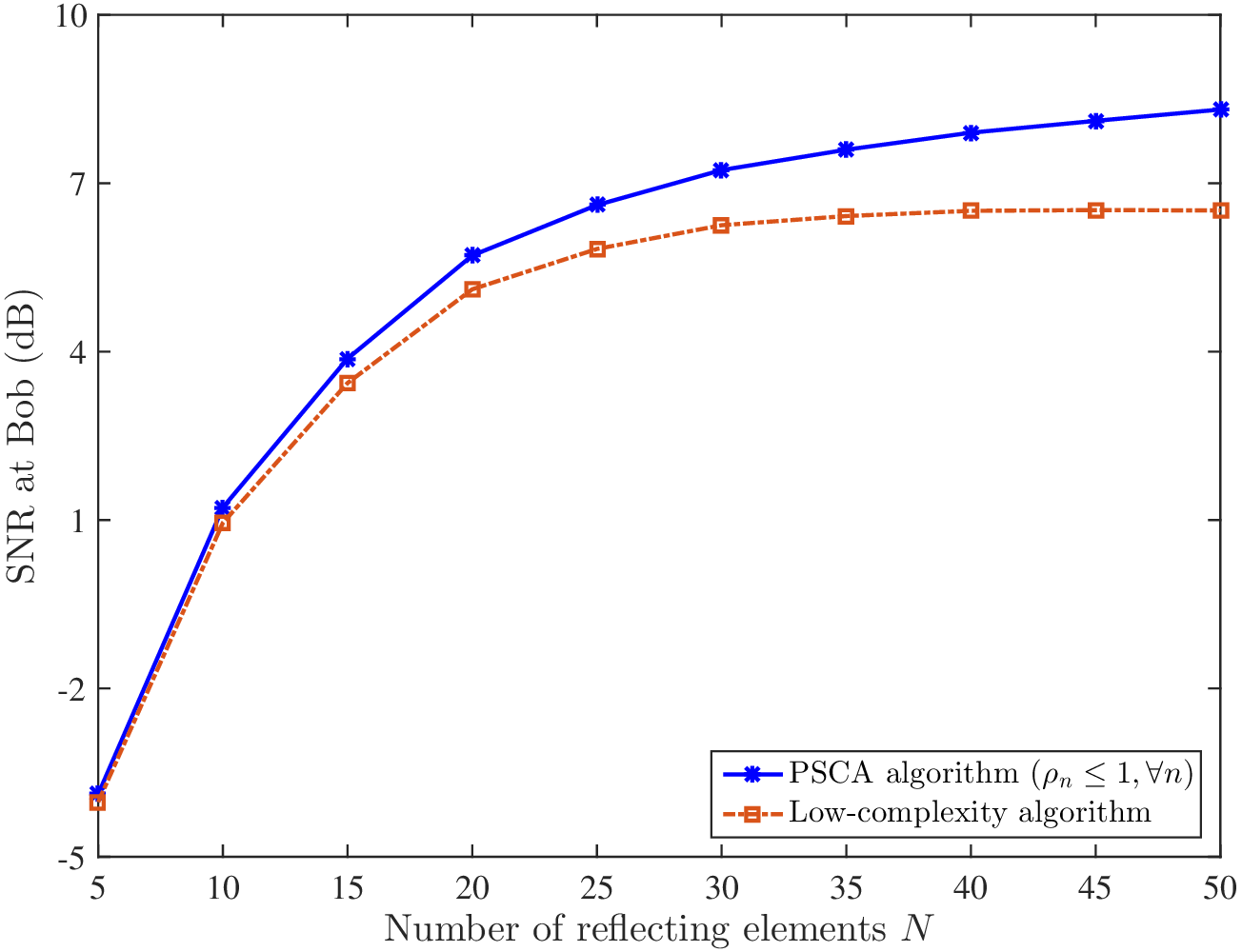}
  \caption{Received SNR at Bob versus the number of reflecting elements at IRS for $\epsilon=0$ (i.e., perfect covertness).}\label{Absolute_covertness}
\end{figure}

In Fig.~\ref{Absolute_covertness}, as expected we observe that the SNR achieved by all the schemes increases with $N$. In addition to the reasons presented in Fig.~2, another reason is that the value of $\sum_{n=1}^N|a_n|$ (i.e., the total channel quality of the reflected path from Alice to Willie) increases with $N$. This leads to the fact that the condition of perfect covertness (i.e., $\sum_{n=1}^N|a_n|\geq |h_{aw}|$) becomes easier to be satisfied as $N$ increases. In general, this figure confirms that perfect covertness can indeed be obtained with the aid of IRS in the context of covert communications, which also verifies the correctness of Theorem~\ref{theorem1}.


\subsection{Without Willie's Instantaneous CSI}

\begin{figure}[!t]
  \centering
  \includegraphics[width=3.4in, height=2.6in]{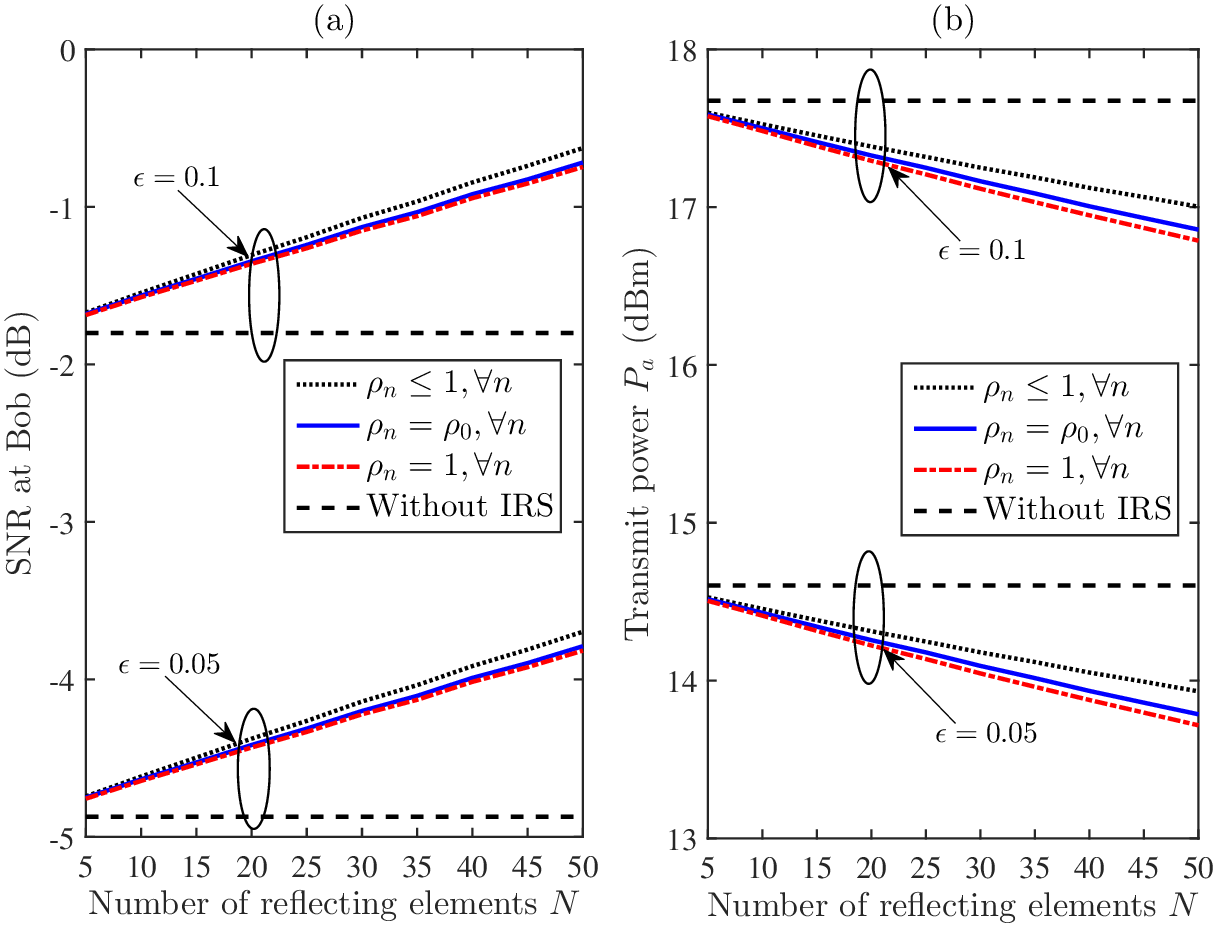}
  \caption{Received SNR at Bob and transmit power at Alice versus the number of reflecting elements for different values of covertness level $\epsilon$, where $\sigma_b^2=-90~\mathrm{dBm}$.}\label{Num_CDI}
\end{figure}

In this subsection, corresponding the considered three different scenarios, ``$\rho_n= 1, \forall n$" denotes the solution with the IRS's reflection amplitude being fixed to $1$, ``$\rho_n=\rho_0, \forall n$" denotes the solution with all elements of the IRS sharing the same reflection amplitude $\rho_0$, and ``$\rho_n\leq 1, \forall n$" denotes solution with the optimal reflection amplitude of each IRS element. We note that the transmit power $P_a$ is optimally designed in the above three solutions.

In Fig.~\ref{Num_CDI}, we plot Bob's SNR and Alice's transmit power $P_a$ achieved by our solutions for different values of $\epsilon$. As expected, we first observe from Fig.~\ref{Num_CDI}(a) that Bob's SNR achieved by all solutions decreases as the covertness level becomes harsher.
We also observe that the solution with ``$\rho_n\leq 1, \forall n$" achieves the highest SNR relative to the solutions of both ``$\rho_n= \rho_0, \forall n$" and ``$\rho_n=1, \forall n$", while the solution with ``$\rho_n= \rho_0, \forall n$" achieves a higher SNR than the solution with ``$\rho_n=1, \forall n$". This demonstrates that the IRS reflection amplitude control can effectively improve the covert communication performance when Willie's instantaneous CSI is not available. In Fig.~\ref{Num_CDI}(b), we first observe that the required transmit power $P_a$ achieved by our developed solutions decreases as the number of reflection elements $N$ increases, which is different from that observed in the case with Willie's instantaneous CSI, where $P_a$ increases with $N$. This is attributed to the fact
 that without Willie's instantaneous CSI, the covertness constraint is only related to the IRS reflection amplitudes and Alice's transmit power $P_a$, but is independent of IRS phase shifts. As such, the average covertness constraint $\frac{P_a}{\sigma_w^2}\left(\chi_{rw}\sum_{n=1}^N\rho_n^2|h_{ar_n}|^2+\chi_{aw}\right)\leq \bar{\epsilon}$ is harder to be satisfied as $N$ increases. Interestingly, we also observe that $P_a$
is higher in the system without IRS relative to that in the system with IRS, which is completely different from the case with Willie's instantaneous CSI.


\begin{figure}[!t]
  \centering
  \includegraphics[width=3.4in, height=2.6in]{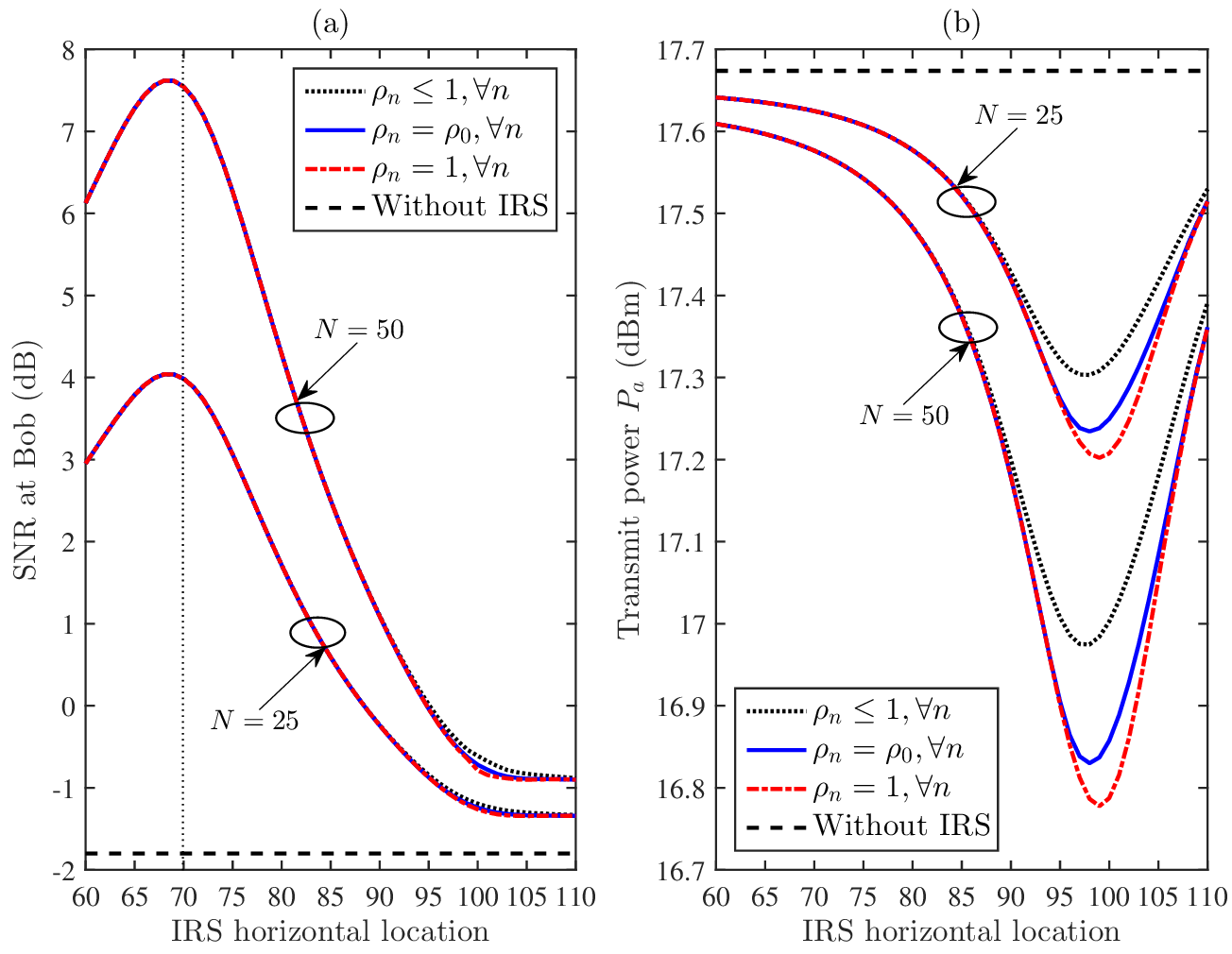}
  \caption{Received SNR at Bob and transmit power at Alice versus IRS horizontal location for covertness level $\epsilon=0.1$, where $\sigma_b^2=-90~\mathrm{dBm}$.}\label{Location_CDI}
\end{figure}

In Fig.~\ref{Location_CDI}, we plot Bob's SNR and Alice's transmit power $P_a$ achieved by different solutions versus the IRS horizontal location. We first observe from Fig.~\ref{Location_CDI}(a) that the optimal IRS horizontal location is close to the LHS of Bob (denoted by the verticle dashed line in this figure). This is due to the fact that satisfying the covertness constraint becomes challenging as IRS moves closer to Willie.
We note that the optimal IRS horizontal location is to balance the transmission quality and communication covertness, since it not only guarantees a small reflection path loss from IRS to Bob, but also makes the covertness constraint easier to be satisfied. We also note that this observation is different from the case with Willie's instantaneous CSI, in which the covertness constraint becomes easier to be satisfied as the IRS moves closer to Willie.
In addition, we observe from Fig.~\ref{Location_CDI}(b) that the transmit power $P_a$ achieved by our developed solutions decreases as the IRS moves closer to Willie.
This is attributed to the fact that the covertness constraint becomes harder to be satisfied as the reflect-path gain from IRS to Willie becomes larger.



\section{Conclusion}


This paper tackled covert communication system designs by considering the assistance of an IRS in the cases with global CSI and without Willie's instantaneous CSI.
 For the case with global CSI, we proved that the perfect covertness can be achieved if the channel quality
of the reflected path is higher than that of the direct path.
Then, we developed a PSCA algorithm and a low-complexity two-stage algorithm to jointly design the IRS's reflection coefficients and Alice's transmit power.
For the case without the Willie's instantaneous CSI, our analysis showed that the phase shift of each IRS element is independent of the covertness constraint, based on which the optimal phase shifts and reflection amplitudes together with the Alice's transmit power are determined. Our examinations showed that deploying an IRS is able to enhance the reflection signal at Bob and deteriorate the detection performance at Willie by properly designing the reflection coefficients, so as to improve covert communication performance. Interestingly, it was revealed  that the optimal horizontal location of the IRS is between Bob and Willie for the case with global CSI, while it is close to the LHS of Bob for the case  without Willie's instantaneous CSI. To further unleash the potential of IRS-aided covert communications,  one challenge to be addressed in future works is how to obtain accurate CSI covertly without using the traditional pilot-based channel estimation methods. Addressing this challenge may call for new emerging techniques (e.g., machine learning) to conduct passive channel estimation (e.g., based on three-dimensional images).


\appendices
\vspace{-0.3cm}
\section{Proof of Theorem \ref{theorem1}}\label{App_the1}

We first note that the perfect covertness implies the required covertness level $\epsilon=0$. In addition, we observe from (P1) that $\mathcal{D}(\mathbb{P}_0|\mathbb{P}_1)$ in covertness constraint \eqref{PF1b} monotonically increases with the received energy at Willie (i.e., $P_a|\mathbf{v}^H\mathbf{a}+h_{aw}|^2$), and $P_a|\mathbf{v}^H\mathbf{a}+h_{aw}|^2=0$ means Willie cannot detect any transmission. Thus, under the perfect covertness constraint, problem (P1) can be reformulated into problem $(\mathrm{P}1')$.
We note that problem $(\mathrm{P}1')$ is not always feasible when the transmit power $P_a$ is non-zero.
In addition, one can verify that $P_a=P_{\max}$ is the optimal solution to problem $(\mathrm{P}1')$ when $E_w\triangleq|\mathbf{v}^H\mathbf{a}+h_{aw}|^2=0$.
In the following, we focus on deriving the condition of $E_w=0$. To this end, we first recall that $v_n=\rho_ne^{-j\theta_n}$, $\forall n$, it follows that
\begin{align}\label{Low_7}
E_w=\left|\sum_{n=1}^N\rho_n|a_n|e^{j\left(\arg(a_n)+\theta_n\right)}+|h_{aw}|e^{j\arg(h_{aw})}\right|^2.
\end{align}
We note that $E_w\!=\!0$ only when the signs of the real and imaginary part of $\sum_{\!n=1\!}^N\rho_n|a_n|e^{j\left(\arg(a_n)\!+\!\theta_n\right)}$ and $|h_{aw}|e^{j\arg(h_{aw})}$ are opposite. We also note that if each summation term $\rho_n|a_n|e^{j\left(\arg(a_n)+\theta_n\right)}$ has the same phase for coherent combining, the synthesized $\sum_{n=1}^N\rho_n|a_n|e^{j\left(\arg(a_n)+\theta_n\right)}$ achieves the largest modulus, which is given by $\sum_{n=1}^N{\rho_n|a_n|}$.
As a result, if
\begin{align}\label{Low_8}
\sum_{n=1}^N\rho_n|a_n|\geq |h_{aw}|,
\end{align}
we can always adjust reflection amplitude $\rho_n$ and reflection phase shift $\theta_n$ such that $E_w=0$. We note that the maximum value of $\sum_{n=1}^N\rho_n|a_n|$ is $\sum_{n=1}^N|a_n|$ due to $\rho_n\in[0,1],\forall n$. Following above discussions, we can conclude that the perfect covertness can be achieved with non-zero transmit power if and only if $\sum_{n=1}^N|a_n|\geq |h_{aw}|$.
This completes the proof of Theorem~\ref{theorem1}.

\vspace{-0.3cm}
\section{Proof of Lemma \ref{lemma1}}\label{App_lem1}

{We first note that the second term on the LHS of \eqref{PCCP_4} is a linear function of the concerned optimization variable $\mathbf{W}$, while the right-hand side (RHS) of \eqref{PCCP_4} is a constant. As such, we only need to prove that the first term on the LHS of \eqref{PCCP_4} is convex with respect to $\mathbf{W}$, which is redefined as
\vspace{-0.05cm}
\begin{align}\label{APPA_1}
 f(\mathbf{W})=\left(1+\frac{\mathrm{Tr}(\mathbf{A}\mathbf{W})}{\sigma_w^2}\right)\ln\left(1+\frac{\mathrm{Tr}(\mathbf{A}\mathbf{W})}{\sigma_w^2}\right).
\end{align}
To proceed, we first note that $x\ln(x)$ is a convex function of $x$ for $x>0$\cite{Boyd}. Then, following the fact that $\mathbf{A}\succeq0$ and $\mathbf{W}\succeq0$, we have that $f(\mathbf{W})$ is convex with respect to $\mathbf{W}$ since the affine transformation of $x\ln(x)$ is convex with respect to $x$. This completes the proof of Lemma~\ref{lemma1}.}
\vspace{-0.4cm}
\section{Proof of Theorem \ref{theorem2}}\label{App_the2}

As per \eqref{Sp2_1}, in order to determine $\mathbb{E}_X\left[\mathcal{D}(\mathbb{P}_0|\mathbb{P}_1)\right]$, we have to derive the distribution of $X$. To this end, we first rewrite $\mathbf{h}^H_{rw}\mathbf{\Theta}\mathbf{h}_{ar}+h_{aw}$ as
\begin{align}\label{Sp2_2}
\mathbf{h}^H_{rw}\mathbf{\Theta}\mathbf{h}_{ar}+h_{aw}=\sum_{n=1}^Nh_{rw_n}^*\rho_ne^{j\theta_n}h_{ar_n}+h_{aw},
\end{align}
where $h_{rw_n}$ and $h_{ar_n}$ are the $n$-th element of $\mathbf{h}_{rw}$ and $\mathbf{h}_{ar}$, respectively. As a result, $\mathbf{h}^H_{rw}\mathbf{\Theta}\mathbf{h}_{ar}+h_{aw}$ follows the distribution $\mathcal{CN}(0,\delta)$, where $\delta\triangleq\chi_{rw}\sum_{n=1}^N\rho_n^2|h_{ar_n}|^2+\chi_{aw}$.
Then, the probability density function (pdf) of $X$ denoted as $f_X(x)$
is an exponential distribution with parameter $\delta^{-1}$. As such, $\mathbb{E}_X\left[\mathcal{D}(\mathbb{P}_0|\mathbb{P}_1)\right]$ can be rewritten as
\begin{align}\label{Sp2_4}
&\mathbb{E}_X\left[\mathcal{D}(\mathbb{P}_0|\mathbb{P}_1)\right]\nonumber\\
&=\int_0^{\infty}L\Bigg[\ln\left(1+\frac{P_ax}{\sigma_w^2}\right)-\frac{P_ax}{P_ax+\sigma_w^2}\Bigg]f_{X}(x)dx,\nonumber\\
&=L\left(1+\frac{\sigma_w^2}{\delta P_a}\right)e^{\frac{\sigma_w^2}{\delta P_a}}E_1\left(\frac{\sigma_w^2}{\delta P_a}\right)-L.
\end{align}
Following \eqref{Sp2_4}, the covertness constraint $\mathbb{E}_X\left[\mathcal{D}(\mathbb{P}_0|\mathbb{P}_1)\right]\leq2\epsilon^2$ can be equivalently rewritten as
\begin{align}\label{Sp2_5}
g\left(\frac{\sigma_w^2}{\delta P_a}\right)\leq 1+ \frac{2\epsilon^2}{L},
\end{align}
where $g\left(\frac{\sigma_w^2}{\delta P_a}\right)\triangleq\left(1+\frac{\sigma_w^2}{\delta P_a}\right)e^{\frac{\sigma_w^2}{\delta P_a}}E_1\left(\frac{\sigma_w^2}{\delta P_a}\right)$.
We note that an exact analytical expression for $g\left(\frac{\sigma_w^2}{\delta P_a}\right)$
is mathematically intractable, since it involves an exponential integral function.

To overcome this difficulty, we next show that $g\left(\frac{\sigma_w^2}{\delta P_a}\right)$ is a monotonically decreasing function of $\frac{\sigma_w^2}{\delta P_a}$. To this end, we first define $g(x)=(1+x)e^xE_1(x)$, where $x\geq 0$. Then, the first derivative of $g(x)$ with respect to $x$ is given by
\begin{align}\label{APPD_1}
\frac{dg(x)}{dx}&=(x+2)e^xE_1(x)-\frac{x+1}{x}\nonumber\\
&=(x+2)e^x\left(\int_x^{\infty}\frac{e^{-t}}{t}dt-s(x)\right),
\end{align}
where $s(x)=\frac{x+1}{x(x+2)e^x}$. We note that the first derivative of $s(x)$ with respect to $x$ is given by
\begin{align}\label{APPD_2}
\frac{ds(x)}{dx}=\frac{e^{-x}\left(-2-x(x+2)^2\right)}{x^2(x+2)^2}.
\end{align}
Considering that $\frac{ds(x)}{dx}$ is a continuous function of $x$ for $x>0$, we have
\begin{align}\label{APPD_3}
\int_x^{\infty}\frac{e^{-t}\left(-2-t(t+2)^2\right)}{t^2(t+2)^2}dt&=s(\infty)-s(x)\nonumber\\
&=-s(x).
\end{align}

As per \eqref{APPD_3}, we can rearrange \eqref{APPD_1} as
\begin{align}\label{APPD_4}
\frac{dg(x)}{dx}&=(x+2)e^x\left(\int_x^{\infty}\frac{e^{-t}}{t}+\frac{e^{-t}\left(-2-t(t+2)^2\right)}{t^2(t+2)^2}dt\right)\nonumber\\
&=(x+2)e^x\int_x^{\infty}\frac{-2e^{-t}}{t^2(t+2)^2}dt.
\end{align}
Following the fact that the integrand $\frac{-2e^{-t}}{t^2(t+2)^2}$ is always less than or equal to $0$, we have $\frac{dg(x)}{dx}\leq 0$ must hold for $x>0$. As a result, $g(x)$ is a monotonically decreasing function of $x$ for $x> 0$.

Following the above fact, we have $g\left(\frac{\sigma_w^2}{\delta P_a}\right)$ given in \eqref{Sp2_5} as a monotonically increasing function of $\frac{\delta P_a}{\sigma_w^2}$. Then, the covertness constraint \eqref{Sp2_5} can be equivalently rewritten as that given in \eqref{Sp2_6}, which completes the proof of Theorem~\ref{theorem2}.

\bibliographystyle{IEEEtran}

\bibliography{IEEEfull,IRS_Covert}

\begin{thebibliography}{10}
\providecommand{\url}[1]{#1}
\csname url@samestyle\endcsname
\providecommand{\newblock}{\relax}
\providecommand{\bibinfo}[2]{#2}
\providecommand{\BIBentrySTDinterwordspacing}{\spaceskip=0pt\relax}
\providecommand{\BIBentryALTinterwordstretchfactor}{4}
\providecommand{\BIBentryALTinterwordspacing}{\spaceskip=\fontdimen2\font plus
\BIBentryALTinterwordstretchfactor\fontdimen3\font minus
  \fontdimen4\font\relax}
\providecommand{\BIBforeignlanguage}[2]{{%
\expandafter\ifx\csname l@#1\endcsname\relax
\typeout{** WARNING: IEEEtran.bst: No hyphenation pattern has been}%
\typeout{** loaded for the language `#1'. Using the pattern for}%
\typeout{** the default language instead.}%
\else
\language=\csname l@#1\endcsname
\fi
#2}}
\providecommand{\BIBdecl}{\relax}
\BIBdecl

\bibitem{Key20175G}
V.~W.~S. Wong, R.~Schober, D.~W.~K. Ng, and L.-C. Wang, \emph{Key technologies
  for {5G} wireless systems}.\hskip 1em plus 0.5em minus 0.4em\relax Cambridge
  U.K.: Cambridge Univ. Press, 2017.

\bibitem{Zhang2020Prospective}
J.~{Zhang}, E.~{Bj\"{o}rnson}, M.~{Matthaiou}, D.~W.~K. {Ng}, H.~{Yang}, and
  D.~J. {Love}, ``Prospective multiple antenna technologies for beyond {5G},''
  \emph{IEEE J. Sel. Areas Commun.}, vol.~38, no.~8, pp. 1637--1660, Jun. 2020.

\bibitem{Wu2021IRStutorial}
Q.~Wu, S.~Zhang, B.~Zheng, C.~You, and R.~Zhang, ``Intelligent reflecting
  surface-aided wireless communications: A tutorial,'' \emph{IEEE Trans.
  Commun.}, vol.~69, no.~5, pp. 3313--3351, May 2021.

\bibitem{Ning2020Beamforming}
B.~{Ning}, Z.~{Chen}, W.~{Chen}, and J.~{Fang}, ``Beamforming optimization for
  intelligent reflecting surface assisted {MIMO}: A sum-path-gain maximization
  approach,'' \emph{IEEE Wireless Commun. Lett.}, vol.~9, no.~7, pp.
  1105--1109, Jul. 2020.

\bibitem{Han2019Large}
Y.~{Han}, W.~{Tang}, S.~{Jin}, C.~{Wen}, and X.~{Ma}, ``Large intelligent
  surface-assisted wireless communication exploiting statistical {CSI},''
  \emph{IEEE Trans. Veh. Technol.}, vol.~68, no.~8, pp. 8238--8242, Aug. 2019.

\bibitem{Peng2021AnalysisRis}
Z.~{Peng}, T.~{Li}, C.~{Pan}, H.~{Ren}, W.~{Xu}, and M.~D. {Renzo}, ``Analysis
  and optimization for {RIS}-aided multi-pair communications relying on
  statistical {CSI},'' \emph{IEEE Trans. Veh. Technol.}, vol.~70, no.~4, pp.
  3897--3901, Apr. 2021.

\bibitem{Zhi2021PowerSca}
K.~Zhi, C.~Pan, H.~Ren, and K.~Wang, ``Power scaling law analysis and phase
  shift optimization of {RIS}-aided massive {MIMO} systems with statistical
  {CSI},'' [Online] Available: https://arxiv.org/abs/2010.13525.

\bibitem{Wu2019Intelligent1}
Q.~{Wu} and R.~{Zhang}, ``Intelligent reflecting surface enhanced wireless
  network via joint active and passive beamforming,'' \emph{IEEE Trans.
  Wireless Commun.}, vol.~18, no.~11, pp. 5394--5409, Mar. 2019.

\bibitem{Huang2019Reconfigurable}
C.~{Huang}, A.~{Zappone}, G.~C. {Alexandropoulos}, M.~{Debbah}, and C.~{Yuen},
  ``Reconfigurable intelligent surfaces for energy efficiency in wireless
  communication,'' \emph{IEEE Trans. Wireless Commun.}, vol.~18, no.~8, pp.
  4157--4170, Aug. 2019.

\bibitem{Pan2020Multicell}
C.~{Pan}, H.~{Ren}, K.~{Wang}, W.~{Xu}, M.~{Elkashlan}, A.~{Nallanathan}, and
  L.~{Hanzo}, ``Multicell {MIMO} communications relying on intelligent
  reflecting surfaces,'' \emph{IEEE Trans. Wireless Commun.}, vol.~19, no.~8,
  pp. 5218--5233, Aug. 2020.

\bibitem{Huang2020Jsac}
C.~{Huang}, R.~{Mo}, and C.~{Yuen}, ``Reconfigurable intelligent surface
  assisted multiuser {MISO} systems exploiting deep reinforcement learning,''
  \emph{IEEE J. Sel. Areas Commun.}, vol.~38, no.~8, pp. 1839--1850, Aug. 2020.

\bibitem{Pan2020Intelligent1}
C.~{Pan}, H.~{Ren}, K.~{Wang}, M.~{Elkashlan}, A.~{Nallanathan}, J.~{Wang}, and
  L.~{Hanzo}, ``Intelligent reflecting surface aided {MIMO} broadcasting for
  simultaneous wireless information and power transfer,'' \emph{IEEE J. Sel.
  Areas Commun.}, vol.~38, no.~8, pp. 1719--1734, Aug. 2020.

\bibitem{Wu2020Weighted}
Q.~{Wu} and R.~{Zhang}, ``Weighted sum power maximization for intelligent
  reflecting surface aided {SWIPT},'' \emph{IEEE Wireless Commun. Lett.},
  vol.~9, no.~5, pp. 586--590, May 2020.

\bibitem{Pan2021Reconfigurable6G}
C.~Pan \emph{et~al.}, ``Reconfigurable intelligent surfaces for {6G} systems:
  Principles, applications, and research directions,'' \emph{IEEE Commun.
  Mag.}, vol.~59, no.~6, pp. 14--20, Jun. 2021.

\bibitem{Huang2020Holog}
C.~{Huang}, S.~{Hu}, G.~C. {Alexandropoulos}, A.~{Zappone}, C.~{Yuen},
  R.~{Zhang}, M.~D. {Renzo}, and M.~{Debbah}, ``Holographic {MIMO} surfaces for
  {6G} wireless networks: Opportunities, challenges, and trends,'' \emph{IEEE
  Wireless Commun.}, vol.~27, no.~5, pp. 118--125, Oct. 2020.

\bibitem{Cui2019Secure}
M.~{Cui}, G.~{Zhang}, and R.~{Zhang}, ``Secure wireless communication via
  intelligent reflecting surface,'' \emph{IEEE Wireless Commun. Lett.}, vol.~8,
  no.~5, pp. 1410--1414, Oct. 2019.

\bibitem{Chu2020Intelligent}
Z.~{Chu}, W.~{Hao}, P.~{Xiao}, and J.~{Shi}, ``Intelligent reflecting surface
  aided multi-antenna secure transmission,'' \emph{IEEE Wireless Commun.
  Lett.}, vol.~9, no.~1, pp. 108--112, Jan. 2020.

\bibitem{Guan2020Intelligent}
X.~{Guan}, Q.~{Wu}, and R.~{Zhang}, ``Intelligent reflecting surface assisted
  secrecy communication: Is artificial noise helpful or not?'' \emph{IEEE
  Wireless Commun. Lett.}, vol.~9, no.~6, pp. 778--782, Jun. 2020.

\bibitem{Dong2020Enhancing}
L.~{Dong} and H.~{Wang}, ``Enhancing secure {MIMO} transmission via intelligent
  reflecting surface,'' \emph{IEEE Trans. Wireless Commun.}, vol.~19, no.~11,
  pp. 7543--7556, Nov. 2020.

\bibitem{Dong2020Secure}
------, ``Secure {MIMO} transmission via intelligent reflecting surface,''
  \emph{IEEE Wireless Commun. Lett.}, vol.~9, no.~6, pp. 787--790, Jun. 2020.

\bibitem{Artificial2020Hong}
S.~{Hong}, C.~{Pan}, H.~{Ren}, K.~{Wang}, and A.~{Nallanathan},
  ``Artificial-noise-aided secure {MIMO} wireless communications via
  intelligent reflecting surface,'' \emph{IEEE Trans. Commun.}, vol.~68,
  no.~12, pp. 7851--7866, Dec. 2020.

\bibitem{Yu2020Robust}
X.~{Yu}, D.~{Xu}, Y.~{Sun}, D.~W.~K. {Ng}, and R.~{Schober}, ``Robust and
  secure wireless communications via intelligent reflecting surfaces,''
  \emph{IEEE J. Sel. Areas Commun.}, vol.~38, no.~11, pp. 2637--2652, Nov.
  2020.

\bibitem{Hong2020Robust}
S.~{Hong}, C.~{Pan}, H.~{Ren}, K.~{Wang}, K.~K. {Chai}, and A.~{Nallanathan},
  ``Robust transmission design for intelligent reflecting surface aided secure
  communication systems with imperfect cascaded {CSI},'' \emph{IEEE Trans.
  Wireless Commun.}, vol.~20, no.~4, pp. 2487--2501, Apr. 2021.

\bibitem{Yan2019Low}
S.~{Yan}, X.~{Zhou}, J.~{Hu}, and S.~V. {Hanly}, ``Low probability of detection
  communication: Opportunities and challenges,'' \emph{IEEE Wireless Commun.},
  vol.~26, no.~5, pp. 19--25, Oct. 2019.

\bibitem{BiaoHe2017on}
B.~He, S.~Yan, X.~Zhou, and V.~K.~N. Lau, ``On covert communication with noise
  uncertainty,'' \emph{IEEE Commun. Lett.}, vol.~21, no.~4, pp. 941--944, Apr.
  2017.

\bibitem{Wang2019Covert}
J.~{Wang}, W.~{Tang}, Q.~{Zhu}, X.~{Li}, H.~{Rao}, and S.~{Li}, ``Covert
  communication with the help of relay and channel uncertainty,'' \emph{IEEE
  Wireless Commun. Lett.}, vol.~8, no.~1, pp. 317--320, Feb. 2019.

\bibitem{Bash2013Limits}
B.~A. Bash, D.~Goeckel, and D.~Towsley, ``Limits of reliable communication with
  low probability of detection on {AWGN} channels,'' \emph{IEEE J. Sel. Areas
  Commun.}, vol.~31, no.~9, pp. 1921--1930, Sep. 2013.

\bibitem{Shahzad2018Achieving}
K.~{Shahzad}, X.~{Zhou}, S.~{Yan}, J.~{Hu}, F.~{Shu}, and J.~{Li}, ``Achieving
  covert wireless communications using a full-duplex receiver,'' \emph{IEEE
  Trans. Wireless Commun.}, vol.~17, no.~12, pp. 8517--8530, Dec. 2018.

\bibitem{Li2020Optimal}
K.~{Li}, P.~A. {Kelly}, and D.~{Goeckel}, ``Optimal power adaptation in covert
  communication with an uninformed jammer,'' \emph{IEEE Trans. Wireless
  Commun.}, vol.~19, no.~5, pp. 3463--3473, May 2020.

\bibitem{Hu2019Covert}
J.~{Hu}, S.~{Yan}, F.~{Shu}, and J.~{Wang}, ``Covert transmission with a
  self-sustained relay,'' \emph{IEEE Trans. Wireless Commun.}, vol.~18, no.~8,
  pp. 4089--4102, Aug. 2019.

\bibitem{Zhou2019Joint}
X.~{Zhou}, S.~{Yan}, J.~{Hu}, J.~{Sun}, J.~{Li}, and F.~{Shu}, ``Joint
  optimization of a {UAV}'s trajectory and transmit power for covert
  communications,'' \emph{IEEE Trans. Signal Process.}, vol.~67, no.~16, pp.
  4276--4290, Aug. 2019.

\bibitem{Wang2020Secrecy}
H.~{Wang}, Y.~{Zhang}, X.~{Zhang}, and Z.~{Li}, ``Secrecy and covert
  communications against {UAV} surveillance via multi-hop networks,''
  \emph{IEEE Trans. Commun.}, vol.~68, no.~1, pp. 389--401, Jan. 2020.

\bibitem{yan2018gaussian}
S.~{Yan}, Y.~{Cong}, S.~V. {Hanly}, and X.~{Zhou}, ``Gaussian signalling for
  covert communications,'' \emph{IEEE Trans. Wireless Commun.}, vol.~18, no.~7,
  pp. 3542--3553, Jul. 2019.

\bibitem{HeB2018Covert}
B.~He, S.~Yan, X.~Zhou, and H.~Jafarkhani, ``Covert wireless communication with
  a poisson field of interferers,'' \emph{IEEE Trans. Wireless Commun.},
  vol.~17, no.~9, pp. 6005--6017, Sep. 2018.

\bibitem{Zheng2019Multi}
T.~{Zheng}, H.~{Wang}, D.~W.~K. {Ng}, and J.~{Yuan}, ``Multi-antenna covert
  communications in random wireless networks,'' \emph{IEEE Trans. Wireless
  Commun.}, vol.~18, no.~3, pp. 1974--1987, Mar. 2019.

\bibitem{Shihao2018Delay}
S.~Yan, B.~He, X.~Zhou, Y.~Cong, and A.~L. Swindlehurst, ``Delay-intolerant
  covert communications with either fixed or random transmit power,''
  \emph{IEEE Trans. Inf. Forensics Security}, vol.~14, no.~1, pp. 129--140,
  Jan. 2019.

\bibitem{Lu2020Intelligent}
X.~{Lu}, E.~{Hossain}, T.~{Shafique}, S.~{Feng}, H.~{Jiang}, and D.~{Niyato},
  ``Intelligent reflecting surface enabled covert communications in wireless
  networks,'' \emph{IEEE Netw.}, vol.~34, no.~5, pp. 148--155, Sep. 2020.

\bibitem{Zan2020Covert}
J.~Si, Z.~Li, J.~Cheng, L.~Guan, and N.~Al-Dhahir, ``Covert transmission
  assisted by intelligent reflecting surface,'' \emph{IEEE Trans. Commun.},
  Early Access, 2021, doi:\textcolor{blue}{\url{10.1109/TCOMM.2021.3082779}}.

\bibitem{Sun2018Short}
X.~{Sun}, S.~{Yan}, N.~{Yang}, Z.~{Ding}, C.~{Shen}, and Z.~{Zhong},
  ``Short-packet downlink transmission with non-orthogonal multiple access,''
  \emph{IEEE Trans. Wireless Commun.}, vol.~17, no.~7, pp. 4550--4564, Jul.
  2018.

\bibitem{Liu2020MatrixCal}
H.~{Liu}, X.~{Yuan}, and Y.~J.~A. {Zhang}, ``Matrix-calibration-based cascaded
  channel estimation for reconfigurable intelligent surface assisted multiuser
  {MIMO},'' \emph{IEEE J. Sel. Areas Commun.}, vol.~38, no.~11, pp. 2621--2636,
  Nov. 2020.

\bibitem{Wei2021ChannelEst}
L.~{Wei}, C.~{Huang}, G.~C. {Alexandropoulos}, C.~{Yuen}, Z.~{Zhang}, and
  M.~{Debbah}, ``Channel estimation for {RIS}-empowered multi-user {MISO}
  wireless communications,'' \emph{IEEE Trans. Commun.}, vol.~69, no.~6, pp.
  4144--4157, Jun. 2021.

\bibitem{Xie2021MaxIRS}
H.~Xie, J.~Xu, and Y.-F. Liu, ``Max-min fairness in {IRS}-aided multi-cell
  {MISO} systems with joint transmit and reflective beamforming,'' \emph{IEEE
  Trans. Wireless Commun.}, vol.~20, no.~2, pp. 1379--1393, Feb. 2021.

\bibitem{MursiaRISMA2021}
P.~Mursia, V.~Sciancalepore, A.~Garcia-Saavedra, L.~Cottatellucci, X.~C.
  P\'{e}rez, and D.~Gesbert, ``{RISMA}: Reconfigurable intelligent surfaces
  enabling beamforming for {IoT} massive access,'' \emph{IEEE J. Sel. Areas
  Commun.}, vol.~39, no.~4, pp. 1072--1085, Apr. 2021.

\bibitem{Abeywickrama2020IRS}
S.~{Abeywickrama}, R.~{Zhang}, Q.~{Wu}, and C.~{Yuen}, ``Intelligent reflecting
  surface: Practical phase shift model and beamforming optimization,''
  \emph{IEEE Trans. Commun.}, vol.~68, no.~9, pp. 5849--5863, Sep. 2020.

\bibitem{Chen2017Survey}
X.~{Chen}, D.~W.~K. {Ng}, W.~H. {Gerstacker}, and H.~{Chen}, ``A survey on
  multiple-antenna techniques for physical layer security,'' \emph{IEEE Commun.
  Surveys Tuts.}, vol.~19, no.~2, pp. 1027--1053, 2nd Quart. 2017.

\bibitem{Boyd}
S.~Boyd and L.~Vandenberghe, \emph{Convex Optimization}.\hskip 1em plus 0.5em
  minus 0.4em\relax Cambridge U.K.: Cambridge Univ. Press, 2004.

\bibitem{Hiriart}
J.~Hiriart-Urruty and C.~Lemarechal, \emph{Convex Analysis and Minimization
  Algorithms I: Fundamentals}.\hskip 1em plus 0.5em minus 0.4em\relax New York:
  Springer, 1996.

\bibitem{Lipp2016}
T.~Lipp and S.~Boyd, ``Variations and extension of the convex-concave
  procedure,'' \emph{Optim. Eng.}, vol.~17, no.~2, pp. 263--287, Jun. 2016.

\bibitem{Li2013Coordinated}
W.~{Li}, T.~{Chang}, C.~{Lin}, and C.~{Chi}, ``Coordinated beamforming for
  multiuser {MISO} interference channel under rate outage constraints,''
  \emph{IEEE Trans. Signal Process.}, vol.~61, no.~5, pp. 1087--1103, Mar.
  2013.

\bibitem{Tervo2015Optimal}
O.~{Tervo}, L.~{Tran}, and M.~{Juntti}, ``Optimal energy-efficient transmit
  beamforming for multi-user {MISO} downlink,'' \emph{IEEE Trans. Signal
  Process.}, vol.~63, no.~20, pp. 5574--5588, Oct. 2015.

\bibitem{Guan2020Joint}
X.~{Guan}, Q.~{Wu}, and R.~{Zhang}, ``Joint power control and passive
  beamforming in {IRS}-assisted spectrum sharing,'' \emph{IEEE Commun. Lett.},
  vol.~24, no.~7, pp. 1553--1557, Jul. 2020.

\end{thebibliography}


\ifCLASSOPTIONcaptionsoff
  \newpage
\fi

\end{document}